\documentclass[%
 reprint,
showpacs,preprintnumbers,
nofootinbib,
 amsmath,amssymb,
 aps,
]{revtex4-1}

\usepackage{graphicx}
\usepackage{dcolumn}
\usepackage{bm}

\usepackage[normalem]{ulem}  
\usepackage[dvips]{color} 

\renewcommand\sout{\bgroup \color{red} \ULdepth=-.5ex \ULset}

\newcommand{\Slash}[1]{\ooalign{\hfil/\hfil\crcr$#1$}}
\newcommand{\re}{\text{Re }}
\newcommand{\im}{\text{Im }}

\newcommand{\bra}[1]{\langle \, #1 \, |}
\newcommand{\ket}[1]{| \, #1 \, \rangle}

\newcommand{\kket}[1]{ \, #1 \, \rangle}

\begin{document}


\title{Compositeness of dynamically generated states in a chiral unitary approach
}

\author{Tetsuo Hyodo}
 \email{hyodo@th.phys.titech.ac.jp}
\affiliation{%
Department of Physics, Tokyo Institute of Technology, 
Meguro 152-8551, Japan
}%

\author{Daisuke Jido}%
\affiliation{%
Yukawa Institute for Theoretical Physics, 
Kyoto University, Kyoto 606--8502, Japan
}%
\affiliation{%
J-PARC Branch, KEK Theory Center,
Institute of Particle and Nuclear Studies, 
High Energy Accelerator Research Organization (KEK),
203-1, Shirakata, Tokai, Ibaraki, 319-1106, Japan
}%

\author{Atsushi Hosaka}%
\affiliation{%
Research Center for Nuclear Physics (RCNP),
Ibaraki, Osaka 567-0047, Japan
}%


\date{\today}

\begin{abstract}
The structure of dynamically generated states in the chiral unitary approach is studied from a viewpoint of their compositeness. We analyze the properties of bound states, virtual states, and resonances in a single-channel chiral unitary approach, paying attention to the energy dependence of the chiral interaction. We define the compositeness of a bound state using the field renormalization constant which is given by the overlap of the bare state and the physical state in the nonrelativistic quantum mechanics, or by the residue of the bound state propagator in the relativistic field theory. The field renormalization constant enables one to define a normalized quantitative measure of compositeness of the bound state. Applying this scheme to the chiral unitary approach, we find that the bound state generated by the energy-independent interaction is always a purely composite particle, while the energy-dependent chiral interaction introduces the elementary component, depending on the value of the cutoff parameter. This feature agrees with the analysis of the effective interaction by changing the cutoff parameter. A purely composite bound state can be realized by the chiral interaction only when the bound state lies at the threshold or when the strength of the two-body attractive interaction is infinitely large. The natural renormalization scheme, introduced by the property of the loop function and the matching with the chiral low-energy theorem, is shown to generate a bound state which is dominated by the composite structure when the binding energy is small.

\end{abstract}


\maketitle


\section{Introduction}

Recently, intensive attention has been paid to the discussion of hadronic molecular states, in which two (or more) hadrons are loosely bound by the interhadron forces. In general, masses of such hadronic molecular states lie close to some two-hadron threshold energies. Classical examples are found in the light quark sector: the $\Lambda(1405)$ resonance below the $\bar{K}N$ threshold~\cite{Dalitz:1959dn,Dalitz:1960du,Dalitz:1967fp} and the scalar mesons $a_0(980)$ and $f_0(980)$ close to the $\bar{K}K$ threshold~\cite{Weinstein:1983gd}. In relation with the recent progress in the hadron spectroscopy at B factories, the charmonium $X(3872)$ resonance is also discussed as the $D\bar{D}^*$ molecule~\cite{Voloshin:2003nt,Tornqvist:2004qy}. It is important to clarify the property and the binding mechanism of the hadronic molecules for the understanding of the nonperturbative aspects of the strong interaction. Various attempts to classify the internal structure of hadrons have been proposed, for instance, by the $N_c$ scaling method~\cite{Pelaez:2003dy,Pelaez:2006nj,Hyodo:2007np,Roca:2008kr,Geng:2008ag,Nagahiro:2011jn}, by property changes along with chiral symmetry restoration~\cite{Hyodo:2010jp}, and by production rates in the heavy-ion collisions~\cite{Cho:2010db,Cho:2011ew}.

The appearance of hadronic molecules near the threshold region is similar to the cluster phenomena in nuclear structure, for instance, the appearance of the $0_2^+$ state of ${}^{12}$C close to the $3\alpha$ threshold (the Hoyle state)~\cite{Funaki:2003af}. For hadrons, however, there are peculiar difficulties in the discussion of the molecular states. For instance, because the $q\bar{q}$ pair creation and annihilation can take place through the strong interaction, ``the number of valence quarks" is not a good classification scheme of the structure. In any case, the discrimination of the hadronic molecules out of other structures is a subtle issue, as compared with the nuclear structure. To start with, we need a theoretical framework with a clear definition of the hadroninc molecular states.

The structure of a particle, whether it is elementary or  composite, was intensively discussed in 1960s, using the field renormalization constant $Z$ in nonrelativistic theories~\cite{Salam:1962ap,Weinberg:1962hj,Weinberg:1963zz,Weinberg:1964zz,PTP29.877}. A famous result is found in the study of the deuteron in Ref.~\cite{Weinberg:1965zz}, where a model-independent relation of the compositeness with observables was derived in the small binding energy limit. The renormalization constant $Z$ is first related to the $p$-$n$-$d$ coupling constant $g$ and the binding energy of the deuteron $B$, which are then connected with the experimental observables such as the scattering length and the effective range. In both steps, small binding energy is assumed. The compositeness through the field renormalization constant $Z$ was also studied in a relativistic field theory~\cite{PR136.B816}, and there have been several attempts to extend the notion of compositeness to resonances~\cite{Morgan:1990ct,Morgan:1992ge,Baru:2003qq,Hanhart:2011jz}.

In this paper, we adopt the strategy to utilize the field renormalization constant for the definition of the compositeness. First, following Ref.~\cite{Weinberg:1965zz}, we evaluate the field renormalization constant $Z$ as the overlap of the bare state in the free Hamiltonian with the physical bound state in the nonrelativistic quantum mechanics. In addition to the universal relation in the weak binding limit~\cite{Weinberg:1965zz}, we also derive the expression in exact form in terms of the scattering amplitude.\footnote{A preliminary discussion on this issue has been given in a conference proceedings~\cite{Hyodo:2010uh}.} Second, following Ref.~\cite{PR136.B816}, we define the field renormalization constant $Z$ as the residue of the full propagator of the bound state in the Yukawa theory. This approach enables us to analyze the energy-dependent interaction. It should be noted that we discuss the compositeness of the bound states in the two-hadron scatterings, regarding the stable hadrons in the scattering states as elementary particles, in the sense that they form the asymptotic states in the nonperturbative QCD vacuum. We do not consider the internal structure of the hadrons in the asymptotic state.

We apply the notion of compositeness to the bound states in a chiral unitary approach~\cite{Kaiser:1995eg,Oset:1998it,Oller:2000fj,Lutz:2001yb,Hyodo:2011ur}. In this model, baryon resonances are described in the multiple scattering of the meson-baryon system whose interaction is given by the chiral low-energy theorem. Although the generated resonance is expected to be a composite of a meson-baryon system, as pointed out in Ref.~\cite{Hyodo:2008xr}, the Castillejo-Dalitz-Dyson (CDD) pole contributions~\cite{Castillejo:1956ed,PR124.264} can be hidden in the cutoff parameter in the renormalization procedure. To avoid this ambiguity, the natural renormalization scheme has been proposed, in which the subtraction constant is chosen to exclude the CDD pole contribution from the loop function. To connect the argument of the compositeness to the natural renormalization scheme, we discuss the properties of the amplitude in the chiral unitary approach in detail.

This paper is organized as follows. In Sec.~\ref{sec:ChU}, we first analyze the properties of the pole singularities in the chiral unitary approach, using a single-channel model. We examine the energy-dependent chiral interaction and its energy-independent counterpart and discuss the properties of the generated states in detail. Next, in Sec.~\ref{sec:compositeness}, we define the compositeness of the bound state using field renormalization constant $Z$ in a general framework of field theories. We consider both nonrelativistic and relativistic approaches, and examine the limit of small binding. The numerical analysis of the compositeness of the bound state in the chiral unitary approach is presented in Sec.~\ref{sec:numerical}.

\section{Chiral unitary approach}\label{sec:ChU}

The meson-baryon scattering and baryon resonances have been well described by the chiral unitary approach, where the chiral low-energy theorem is combined with the unitarity condition of the scattering amplitude~\cite{Kaiser:1995eg,Oset:1998it,Oller:2000fj,Lutz:2001yb}. A detailed formulation of the chiral unitary approach can be found in Ref.~\cite{Hyodo:2011ur}. In this section we construct the single-channel version of the chiral unitary approach along the same line with Ref.~\cite{Hyodo:2008xr} and study the properties of the generated states in the amplitude. It is our aim to establish a tractable model which will be used to examine the argument of the  compositeness in Sec.~\ref{sec:compositeness}. While we adopt the meson-baryon scattering as an example, the scattering of the Nambu-Goldstone (NG) boson with a heavy meson can be described in the same formulation (see Appendix C in Ref.~\cite{Hyodo:2006kg}). The framework of the scattering of the NG bosons are the essentially the same, thanks to the chiral low energy theorem.

\subsection{Construction of the scattering amplitude}

We consider a single-channel $s$-wave scattering of a baryon with mass $M$ and a NG boson with mass $m$. The chiral low energy theorem~\cite{Weinberg:1966kf,Tomozawa:1966jm} dictates the leading interaction proportional to the boson energy $\omega\simeq W-M$ in the following form:
\begin{align}
    V^{\text{(WT)}}(W)
    =& C(W-M) 
     \label{eq:WTinteraction} ,
\end{align}
where $C$ is a real-valued coupling constant with mass dimension minus two and $W$ is the total center-of-mass energy.\footnote{We may recover the usual WT interaction in the chiral unitary approach by replacing $C\to -C_g/(2f^2)$ with the pion decay constant $f$ and the group theoretical factor $C_g$.} Although the coupling constant is determined by the flavor content of the scattering system, here we simply regard it as a parameter of the model. The energy dependence of the interaction is a consequence of the low energy theorem~\cite{Adler:1964um}. To appreciate the significance of the energy dependence, we also consider an energy-independent constant interaction whose strength is normalized to the threshold value of the WT interaction:
\begin{align}
    V^{\text{(const)}}=
    Cm = 
    V^{\text{(WT)}}(W=M+m)
    \label{eq:Eindepinteraction} .
\end{align}
This interaction generates the same amplitude as the WT interaction only at the threshold, and the deviation becomes large in the energy region far from the threshold. In relation with the pole structure of the $\Lambda(1405)$ resonance~\cite{Jido:2003cb,Hyodo:2007jq}, the energy dependence of the chiral interaction has been recently studied~\cite{Ikeda:2010tk,Ikeda:2011dx}.

The scattering amplitude can be obtained by solving the Bethe-Salpeter equation
\begin{align}
    T(W)
    &=V(W) + V(W)G(W)T(W)
    \label{eq:BS} ,
\end{align}
with $V(W)$ being $V^{\text{(WT)}}(W)$ or $V^{\text{(const)}}$. The loop function $G(W)$ is given by
\begin{align}
    G(W)
    =i\int\frac{d^{4}q}{(2\pi)^{4}}
    \frac{2M}{(P-q)^{2}-M^{2}+i\epsilon}
    \frac{1}{q^{2}-m^{2}+i\epsilon}
	\label{eq:Gunrenormalized}
    ,
\end{align}
where $P=(W,\bm{0})$. This integral diverges logarithmically. With the dimensional regularization, we obtain
\begin{align}
    G(W;a)=&\frac{2M}{(4\pi)^{2}}
    \Biggl(a
    +\frac{m^{2}-M^{2}+W^2}{2W^2}\ln\frac{m^{2}}{M^{2}}
    \nonumber\\
    &+\frac{\bar{q}(W)}{W}
    \Bigl\{\ln[W^2-(M^{2}-m^{2})+2W\bar{q}(W)] \nonumber \\
    &+\ln[W^2+(M^{2}-m^{2})+2W\bar{q}(W)] 
    \nonumber\\
    &
    -\ln[-W^2+(M^{2}-m^{2})+2W\bar{q}(W)]\nonumber \\
    &-\ln[-W^2-(M^{2}-m^{2})+2W\bar{q}(W)]
    \Bigr\}\Biggr) ,
    \label{eq:Gfn}
\end{align}
where $a$ is the subtraction constant at the subtraction point $\mu_{s}=M$, which plays a role of the cutoff parameter of the loop function. The three-momentum variable is defined as
\begin{align}
    \bar{q}(W)
    =&\frac{\sqrt{[W^2-(M-m)^2][W^2-(M+m)^2]}}{2W}
    \label{eq:barq} .
\end{align}
$G(W;a)$ is an analytic function of $W$ in the whole complex plane, except for on the real axis above the threshold where it has a branch cut. In conventional approaches, the subtraction constant has been used to fit the experimental data. In Ref.~\cite{Hyodo:2008xr}, on the other hand, the natural value of the subtraction constant $a_{\text{natural}}$ was introduced by the theoretical argument to exclude the CDD pole contribution from the loop function. We discuss the influence of the subtraction constants to the scattering amplitude. A solution of Eq.~\eqref{eq:BS} is given by
\begin{align}
    T(W)
    &=
    \frac{1}{1-V(W)G(W;a)}V(W)
    \label{eq:amplitude} ,
\end{align}
with the help of the on-shell factorization~\cite{Oller:1997ti,Oset:1998it}. The same form of the amplitude can also be obtained by the $N/D$ method with the $N=1$ prescription~\cite{Oller:2000fj}. For the energy-independent interaction~\eqref{eq:Eindepinteraction}, no factorization is needed so that both the $N=1$ prescription and the $N=V$ prescription in the $N/D$ method provide the equivalent amplitude~\eqref{eq:amplitude}. Note that Eq.~\eqref{eq:amplitude} is obtained by neglecting the unphysical (left-hand and circular) cuts, although the unitarity cut is properly taken into account. Thus, this framework should not be applied to the system with a deeply bound state, which might be influenced by the unphysical cuts.

If the interaction is sufficiently attractive, bound states and resonances can be dynamically generated. A bound state appears as a pole of the full amplitude $T(W)$ on the real axis in the first Riemann sheet below the threshold, while a resonance is represented by a pole in the second Riemann sheet of the complex energy plane above the threshold. A pole on the second Riemann sheet below the threshold is called a virtual state, which is a peculiar phenomenon in the $s$-wave scattering. The amplitude in the second Riemann sheet is given by
\begin{align}
    T_{II}(W)
    =&
    \frac{1}{1-V(W)G_{II}(W;a)}V(W), 
    \label{eq:amplitudeII} \\ 
    G_{II}(W;a)
    =&G(W;a)+i\frac{2M\bar{q}(W)}{4\pi W} .
    \label{eq:GfnII}
\end{align}
Note that below the threshold $W<M+m$, $\bar{q}(W)$ is purely imaginary. Therefore, on the real axis below the threshold, the loop function is real in both the first and the second Riemann sheets. For later reference, we point out that $G(W;a)$ [$G^{{II}}(W;a)$] is monotonically decreasing (increasing) function of $W$ below the threshold.

Because the interactions in Eqs.~\eqref{eq:WTinteraction} and \eqref{eq:Eindepinteraction} have no singularity in the complex energy plane except for $|W|\to \infty$, we deduce the condition for the pole of the amplitude as 
\begin{align}
    \begin{cases}
    1-V(z_R)G(z_R;a)
    =
    0 
    &\text{(bound state)} , \\
    1-V(z_R)G_{II}(z_R;a)
    =
    0
    &\text{(virtual state/resonance)} ,
    \end{cases}
    \label{eq:polecond} 
\end{align}
where $z_R$ is the energy of the bound state, virtual state, or resonance. The pole position of $z_R$ should be on the real axis in the first Riemann sheet, while $z_R$ can be complex in the second Riemann sheet. Around the pole at $W=z_R$ we can formally expand the amplitude in the Laurent series as
\begin{align}
    T(W)
    =
    \frac{g^2}{W-z_R}
    +\sum_{n=0}^{\infty}T^{(n)}(W-z_R)^n
    \label{eq:poleamp} ,
\end{align}
with constant coefficients $T^{(n)}$. The first term can be interpreted as the propagator of a particle with mass $z_R$, which couples to the scattering states through the coupling constant $g$. Hence, we identify the residue of the pole in the amplitude $T(W)$ as the energy-independent coupling strength $g^2$.

\subsection{Bound state with constant interaction}\label{subsec:boundindep}

Let us first consider the case when the system develops a bound state with energy-independent interaction~\eqref{eq:Eindepinteraction}. The amplitude has a pole on the real axis in the first Riemann sheet: $ z_R\equiv M_B$, $\im M_B=0$, and $M\leq M_B\leq M+m$ where $M_{B}=M$ is the lower limit of the mass of the bound state; otherwise, the system becomes unstable. The pole condition~\eqref{eq:polecond} leads to
\begin{align}
    1-CmG(M_B;a)
    =0 .
    \nonumber
\end{align}
With this relation, we can determine the value of $C$ such that a bound state is generated at $W=M_B$ with the subtraction constant $a$ as
\begin{align}
    C(M_B;a)
    =\frac{1}{mG(M_B;a)} .
    \label{eq:Eindepstrength}
\end{align}
Here we have indicated that $C$ can be also regarded as a function of $M_{B}$ and $a$, when the bound state is generated. The loop function $G$ is real below the threshold, so it is consistent with the real-valued coupling strength $C$. If $M_B$ were complex, then the loop function $G(M_B)$ would be complex and no solution would be found for Eq.~\eqref{eq:Eindepstrength}. This means that there is no pole in the first Riemann sheet except for on the real axis, in accordance with the causality.

Note that Eq.~\eqref{eq:Eindepstrength} is the relation among $C$, $M_B$, and $a$. In general, the parameters in this model are the strength of the interaction $C$ and the subtraction constant $a$. When there is a bound state, the system can also be characterized by the mass of the bound state $M_B$ and $a$, with the help of the relation~\eqref{eq:Eindepstrength}. Using Eq.~\eqref{eq:poleamp}, we can calculate the coupling constant $g$ from the residue of the amplitude $T$ as~\cite{Gamermann:2009uq}
\begin{align}
    [g(M_B)]^2
    =& \lim_{W\to M_B}(W-M_B)T(W) \nonumber \\
    =&-\left.\frac{1}
    {\frac{\partial G(W;a)}{\partial W}
    }\right|_{W= M_B}
    \equiv
    -
    \frac{1}
    {G^{\prime}(M_B)} .
    \label{eq:Eindepcoupling}
\end{align}
Below the threshold, $G(W;a)$ is real and monotonically decreasing function of $W$~\cite{Hyodo:2008xr}, so we have $G^{\prime}(M_B)< 0$. This indicates that the coupling square is always real and positive:
\begin{align}
    [g(M_B)]^2
    \geq &0 ,
    \nonumber
\end{align}
which leads to the real-valued coupling constant. The equality holds for $M_{B}=M+m$ where $G^{\prime}(M_{B})\to -\infty$. Note also that the dependence on the subtraction constant $a$ vanishes after taking the derivative in Eq.~\eqref{eq:Eindepcoupling}, because the subtraction constant does not depend on the energy.\footnote{In the present framework, we perform the single subtraction to tame the divergence in Eq.~\eqref{eq:Gfn}. We can introduce further subtractions, which results in the energy-dependent subtraction terms. In this case, the coupling constant depends on the subtraction constants. We  consider the case with the general interaction and the general subtraction terms in Appendix~\ref{sec:generalcase}.} Namely, the coupling of the bound state to the scattering state $g$ is unambiguously determined by the mass of the bound state $M_B$ and independent of the subtraction constant $a$. Therefore, through the relation~\eqref{eq:Eindepstrength}, we can freely change the subtraction constant $a$ and the coupling strength $C$, without altering the property of the bound state. In fact, this is one of the consequences of the invariance of the amplitude $T(W)$ under the transformation
\begin{align}
    a\to 
    &a-\Delta a,
     \quad
    C
    \to 
    \left(\frac{1}{C}-\frac{2M}{(4\pi)^2}\Delta a m\right)^{-1}
    \label{eq:Eindeptrans} .
\end{align}
Because the whole amplitude $T(W)$ is invariant, the property of the bound state remains unchanged.

\subsection{Bound state with WT interaction}
\label{subsec:bounddep}

Next we deal with the energy-dependent WT interaction~\eqref{eq:WTinteraction}. The condition for the bound state~\eqref{eq:polecond} becomes
\begin{align}
    1-C(M_B-M)G(M_B;a)
    =0 .
    \label{eq:WTcond}
\end{align}
The coupling strength $g$ can also be calculated from the residue of the pole as
\begin{align}
    [g(M_B;a)]^2 
    =& \lim_{W\to M_B}(W-M_B)T(W)
    \nonumber \\
    =&-
    \frac{1}
    {
    G^{\prime}(M_B)+\frac{G(M_B;a)}{M_B-M}
    } 
    \label{eq:WTcoupling} 
\end{align}
In this case, the coupling constant depends on the subtraction constant $a$, in contrast to Eq.~\eqref{eq:Eindepcoupling}. This is closely related to the energy dependence of the interaction, because Eq.~\eqref{eq:WTcoupling} is obtained by differentiating both the numerator and the denominator, and the energy dependence of the interaction leaves the $G(M_B;a)$ term in the final expression. Note also that if $G(M_B;a)=0$, the result coincides with that of the energy-independent interaction~\eqref{eq:Eindepcoupling}, which, however, means $C\to -\infty$ because of the bound state condition \eqref{eq:WTcond}. This limit will be important in the discussion of the compositeness.

As before, the condition~\eqref{eq:WTcond} relates $C$, $M_B$, and $a$, so we can characterize the system by ($M_B$, $a$) instead of $(C,a)$, keeping the mass of the bound state unchanged. In this case, the interaction strength $C$ such that the bound state appears at $M_B$ is calculated as
\begin{align}
    C(M_B;a) = \frac{1}{(M_B-M)G(M_B;a)} .
    \label{eq:WTstrength}
\end{align}
The state with zero binding energy corresponds to the upper limit for the mass of the bound state $M_B=M+m$, which is called \textit{zero energy resonance}~\cite{Taylor}. The coupling strength for such state, together with the natural value of the subtraction constant~\cite{Hyodo:2008xr} (the natural subtraction constant $a_{\text{natural}}$ is also explained in Appendix~\ref{sec:generalcase}), is the critical coupling introduced in Refs.~\cite{Hyodo:2006yk,Hyodo:2006kg}:
\begin{align}
    C_{\text{crit}} 
    = C(M+m;a_{\text{natural}}) 
    =\frac{1}{mG(M+m;a_{\text{natural}})},
    \nonumber
\end{align}
which is the smallest coupling strength to generate a bound state. 

As discussed in Ref.~\cite{Hyodo:2008xr}, the variation of the subtraction constant can be absorbed by the introduction of a pole term in the interaction kernel, keeping the amplitude $T(W)$ invariant:
\begin{align}
    a\to 
    &a-\Delta a,
    \nonumber
    \\
    V(W)
    \to
    \tilde{V}(W)=
    & C(W-M)
    -C\frac{(W-M)^2}{(W-M_{\text{eff}})} ,
    \label{eq:poleterm} \\
    &M_{\text{eff}}
    =M+\frac{(4\pi)^2}{2M C \Delta a } .
    \label{eq:Meff}
\end{align}
Equation~\eqref{eq:poleterm} shows that the existence of a pole at $W=M_{\text{eff}}$ in the interaction kernel $\tilde{V}(W)$. The residue of this pole is calculated as
\begin{align}
    \lim_{W\to M_{\text{eff}}}(W-M_{\text{eff}})\tilde{V}(W)
    =&-\frac{(4\pi)^4}{4M^2 C (\Delta a)^2 } .
    \label{eq:Meffresidue}
\end{align}
Both the mass and the residue are determined by $\Delta a$. This means that it is not possible to introduce an arbitrary pole term by the variation of the subtraction constant; the mass and the coupling of the pole term are related to each other through Eqs.~\eqref{eq:Meff} and \eqref{eq:Meffresidue}.

This pole term cannot be absorbed by simple shift of the coupling strength of the interaction kernel as in Eq.~\eqref{eq:Eindeptrans}. Actually, Eq.~\eqref{eq:poleterm} can be written in the similar form to Eq.~\eqref{eq:Eindeptrans} as
\begin{align}
    a\to 
    &a-\Delta a,
    \quad
    C
    \to 
    \left(\frac{1}{C}-\frac{2M}{(4\pi)^2}\Delta a(W-M)\right)^{-1}
    .
    \label{eq:WTtrans}
\end{align}
This is the transformation which leaves the amplitude $T(W)$ invariant for the WT interaction. In the case of the energy-independent interaction, the change of the subtraction constant can be compensated by the simple shift of the coupling strength $C$ [Eq.~\eqref{eq:Eindeptrans}], and the coupling constant of the bound state to the scattering state $g$ is independent of the subtraction constant [Eq.~\eqref{eq:Eindepcoupling}]. In contrast, for the WT interaction, the change of the subtraction constant causes the energy-dependent transformation [Eq.~\eqref{eq:WTtrans}] and the coupling constant $g$ depends on the subtraction constant [Eq.~\eqref{eq:WTcoupling}]. These are the important consequences of the energy dependence of the WT interaction. As we see below, this difference is crucial to the structure of the bound state.

Let us consider the properties of the coupling square in Eq.~\eqref{eq:WTcoupling} which must be real and positive to obtain a real-valued coupling constant. Because the mass of the bound state should be larger than the mass of the constituent, we have a condition
\begin{align}
    M_B-M
    \geq &0 .
    \nonumber
\end{align}
It follows from Eq.~\eqref{eq:WTstrength} that $G(M_B;a)\leq 0$ for an attractive interaction $C<0$ (equality holds for $C\to -\infty$). Therefore, $G(M_B;a)/(M_B-M)\leq 0$ and we also have $G^{\prime}(M_B)< 0$. Thus, from Eq.~\eqref{eq:WTcoupling}, for the bound state generated by the attractive WT interaction, we find that the coupling square is always real and positive:
\begin{align}
    [g(M_B;a)]^2
    \geq  &0 .
    \nonumber
\end{align} 
Equality holds for $M_{B}=M+m$, where $G^{\prime}(M_{B})\to -\infty$ and $a$ dependence of the coupling constant vanishes in this limit.

\subsection{Virtual states and resonances}\label{subsec:virtual}

Finally, we consider the case with the pole on the complex energy plane at $W=z_R$ in the second Riemann sheet. If the real part of the pole is higher (lower) than the threshold, the state is a resonance (virtual state).

For the energy-independent interaction, the pole condition~\eqref{eq:polecond} becomes
\begin{align}
    G_{II}(z_R;a)
    =&\frac{1}{Cm} .
    \nonumber
\end{align}
The right-hand side is real, while the left-hand side is, in general, complex. Therefore, at the pole position $W=z_R$ the imaginary part of the loop function should vanish: 
\begin{align}
    \im[G_{II}(z_R;a)] 
    =&0 .
    \label{eq:condition}
\end{align}
Because the subtraction constant $a$ is real and additively included in the loop function $G_{II}(z_R;a)$, this condition is independent of $a$ and is the necessary condition for $z_R$ to be the pole position. We first consider the virtual state on the real axis, $z_{R}=M_{V}$, $\im M_{V}=0$, and $M\leq M_{V}< M+m$. Below the threshold $M_V< M+m$, there is no discontinuity, so the imaginary part of the $G_{II}(M_V)$ function vanishes and the condition~\eqref{eq:condition} is satisfied. This means that a virtual state can be produced on the real axis for a weakly attractive interaction. 

Resonances are associated with the pole $z_{R}$ with a finite imaginary part. As shown in Appendix~\ref{sec:pole}, the pole trajectory is continuous in the complex energy plane as the coupling strength $C$ is varied with a fixed the subtraction constant $a$. Thus, the resonance solution should be connected to the virtual state solution on the real axis at some energy. To have a resonance at $z_R$ with $\im z_R\neq 0$, Eq.~\eqref{eq:condition} should be satisfied in the vicinity of the real axis in the region $M\leq \re z_R< M+m$. This can be implemented if the derivative of the imaginary part in the imaginary direction is zero. However, Cauchy-Riemann's theorem leads to
\begin{align}
    &\left. \frac{\partial }{\partial (\im z_R)}
    \im G_{II}(z_R;a)\right|_{\im z_R=0}
    =\frac{\partial  G_{II}(\re z_R;a)}{\partial (\re z_R)} 
    > 0 ,
    \nonumber
\end{align}
for $M\leq \re z_R< M+m$. Thus, the existence of the branch of the pole trajectory to the resonance solution is not allowed. This means that the pole with a finite imaginary part cannot appear for the energy-independent interaction. This is consistent with the interpretation in the nonrelativistic potential model; simple attractive $s$-wave potential has no centrifugal barrier and thus does not produce a resonance state. 

The residue $g^2$ can be obtained by the derivative of the loop function
\begin{align}
    [g(M_V)]^2
    =&-
    \frac{1}
    {G^{\prime}_{II}(M_V)}  .
    \nonumber
\end{align}
For the virtual state on the real axis $\im M_V=0$, the derivative of the loop function on the second Riemann sheet is positive. Therefore, the coupling square is negative. Because the pole singularities in the second Riemann sheet are interpreted as Gamow states, the coupling constant becomes pure imaginary. As in the case of the bound state with the constant interaction, the dependence of the subtraction constant is lost by taking the derivative.

For the WT interaction, the bound-state condition is
\begin{align}
    1-C(z_R-M)G_{II}(z_R;a)
    =0 .
    \label{eq:WTstrengthR}
\end{align}
Again, $C$ is a real number, so the equation to be satisfied at $z_R$ is
\begin{align}
    \im[(z_R-M)G_{II}(z_R;a)]
    =&0 .
    \label{eq:conditionWT}
\end{align}
In the same way as in the energy-independent interaction, this condition is satisfied on the real axis $z_{R}=M_{V}$, so the virtual state can be formed on the real axis. As for the resonance, in the present case, 
\begin{align}
    &\left.\frac{\partial }{\partial (\im z_R)}
    \im [(z_R-M)G_{II}(z_R;a)]
    \right|_{\im z_R=0} \nonumber \\
    =&    \frac{\partial }{\partial (\re z_R)}
    [(\re z_R-M) G_{II}(\re z_R;a)] \nonumber \\
    =&   
    G_{II}(\re z_R;a)
    +(\re z_R-M) G_{II}^{\prime}(\re z_R) .
    \label{eq:meet}
\end{align}
This can be zero for $M\leq \re z_R< M+m$, if the subtraction constant is properly chosen, because $G_{II}(\re z_R;a)\leq 0$ from Eq.~\eqref{eq:WTstrengthR} and $(\re z_R-M) G_{II}^{\prime}(\re z_R)\geq 0$. Namely, for the energy-dependent interaction, the pole can have the imaginary part and eventually an $s$-wave resonance can be produced. Several examples of the $s$-wave single-channel resonance can be found in hadron physics, such as the $\sigma$ meson in the $\pi\pi$ scattering~\cite{Hyodo:2010jp} and the lower energy pole of $\Lambda(1405)$~\cite{Jido:2003cb,Hyodo:2007jq,Ikeda:2011dx}. 

The condition~\eqref{eq:conditionWT} determines the relation between $\re [z_R]$ and $\im [z_R]$, which depends on the subtraction constant. Namely, for a given $a$, the pole trajectory  is specified in the complex energy plane as a function of $C$ (for illustration, we show the trajectory of the pole position in Appendix~\ref{sec:pole}). The position of the pole is then determined by the coupling strength $C$ and vice versa.

The residue of the pole is
\begin{align}
    [g(z_R;a)]^2
    =&-
    \frac{1}
    {
    G^{\prime}_{II}(z_R)
    +\frac{G_{II}(z_R;a)}{z_R-M}
    } .
    \label{eq:couplingWT}
\end{align}
For a virtual state ($\im z_{R}=0$), the coupling square is given by the real number. On the real axis, $G^{\prime}_{II}(z_{R})$ and $z_{R}-M$ are positive, but the sign of the $G_{II}(z_R;a)$ depends on the value of the subtraction constant. Thus, the sign of the coupling square depends on the subtraction constant and is determined by the balance of the first term and the second term in the denominator of Eq.~\eqref{eq:couplingWT}. Because $G^{\prime}_{II}(z_{R})$ diverges at the threshold, the coupling square is negative for the virtual state close to the threshold. For a resonance ($\im z_{R}\neq 0$), the coupling constant is complex because of $G_{II}$, $G_{II}^{\prime}$ and $(z_R-M)$ which are complex in the complex energy plane. 

\section{Field renormalization constant and compositeness}\label{sec:compositeness}

In this section, we define the compositeness of the bound state in the scattering amplitude in the nonrelativistic quantum mechanics~\cite{Weinberg:1965zz} and in the relativistic field theory~\cite{PR136.B816}. For this purpose, we describe the scattering system with one bound state by the field theory whose free spectrum contains one bare state and two-body scattering states. We define the field renormalization constant $Z$ as the overlap of the physical bound state and the bare state, which characterizes the elementarity of the bound state. Expressing the constant $Z$ in terms of physical quantities, we obtain a master formula of the compositeness. With this master formula, we can discuss the compositeness of bound states, either those observed in experiments or those obtained in a theory such as the chiral unitary approach. To be specific, we consider the $s$-wave scattering of a baryon $\psi(J^{P}=1/2^{+})$ and a meson $\phi(0^{-})$ in which a bound state $B(1/2^{-})$ appears. It is easy to derive similar expressions of the compositeness for other scattering systems with an $s$-wave bound state (see Appendix~\ref{sec:scalar} and Ref.~\cite{PR136.B816}).

\subsection{Compositeness in nonrelativistic quantum mechanics}
\label{subsec:nonrel}

Let us first consider the two-body scattering system with one bound state in a nonrelativistic quantum mechanics. The Hamiltonian of the system $H$ is decomposed into the free Hamiltonian $H_0$ and the interaction $V$:
\begin{align}
    H = H_0 + V .
    \nonumber
\end{align}
We assume that the eigenstates of the free Hamiltonian $H_0$ are given by the continuum states labeled by the relative momentum $\ket{\bm{q}}$ and an elementary state\footnote{In general, we may have several bound states as well as multiple scattering channels (see Appendix~\ref{sec:coupledchannel}).} $\ket{B_0}$ which is orthogonal to $\ket{\bm{q}}$. $V$ stands for the interaction among scattering states, as well as the coupling of the scattering state to the elementary state. $\ket{B_0}$ is elementary in the sense that it is an eigenstate of the Hamiltonian $H_{0}$ and the origin of $\ket{B_0}$ should be attributed to dynamics other than the interaction $V$.\footnote{One may regard $\ket{B_{0}}$ as a bound state formed by another Hamiltonian $H^{\prime}$. Using the Feshbach projection formalism with single bound state approximation~\cite{Feshbach:1958nx,Feshbach:1962ut}, we can set up this situation. Thus, strictly speaking, $\ket{B_{0}}$ is not always an elementary component; it should be considered as a CDD pole contribution which does not originate in the present model space of the scattering. In this paper, however, we use the word ``elementary'' for simplicity.}
The eigenenergies are given by
\begin{align}
    H_0\ket{\bm{q}}
    & = E(q)\ket{\bm{q}} \quad
    H_0\ket{B_0}
    = -B_0\ket{B_0} ,
    \nonumber
\end{align}
where $E(q)=q^2/(2\mu)$ with $\mu$ being the reduced mass of the system. $\ket{\bm{q}}$ and $\ket{B_0}$ are orthogonal and form the complete set:
\begin{align}
    \bra{\bm{q}}\kket{\bm{q}^{\prime}}
    &=\delta(\bm{q}^{\prime}-\bm{q}),
    \quad 
    \bra{B_0}\kket{B_0}=1 , \nonumber \\
    \bra{B_0}\kket{\bm{q}}
    &=0 ,\nonumber
    \\
    1 
    &= \ket{B_0}\bra{B_0} + \int d\bm{q} \ket{\bm{q}}\bra{\bm{q}}
    .
    \label{eq:complete}
\end{align}
We consider that the Hilbert space of the full Hamiltonian also contains one bound state $\ket{B}$ with the binding energy defined as $B>0$. Namely, $\ket{B}$ is the eigenstate of the full Hamiltonian
\begin{align}
    H\ket{B}
    & = (H_0+V)\ket{B} 
    = -B\ket{B} ,\label{eq:Schroedinger} 
\end{align}
where $\ket{B}$ is normalized and orthogonal to the scattering states $\ket{\bm{q},\text{full}}$ as
\begin{align}
    \bra{\bm{q},\text{full}}\kket{\bm{q}^{\prime},\text{full}}
    &=\delta(\bm{q}^{\prime}-\bm{q}),
    \quad 
    \bra{B}\kket{B}=1 , \nonumber \\
    \bra{B}\kket{\bm{q},\text{full}}
    &=0 ,
    \nonumber 
\end{align}
and they form the complete set as
\begin{align}    
    1 
    &= \ket{B}\bra{B} 
    + \int d\bm{q} \ket{\bm{q},\text{full}}
    \bra{\bm{q},\text{full}}
    .
    \label{eq:completefull}
\end{align}
We note that the interaction $V$ should be energy independent, to ensure the orthogonality and the completeness of the Hilbert space of the full Hamiltonian. 

We then define the field renormalization constant $Z_{NR}$ as the probability of finding the bound state $B$ in the bare state $B_0$:
\begin{align}
    Z_{NR}
    & \equiv |\bra{B_0}B\, \rangle |^2  .
    \nonumber
\end{align}
This expresses the elementarity of the bound state $\ket{B}$. It is clear from the completeness~\eqref{eq:complete} that $0\leq Z_{NR}\leq 1$. If $\ket{B}$ is a purely composite (elementary) particle, we obtain $Z_{NR}=0$ ($Z_{NR}=1$). Therefore, $1-Z_{NR}$ serves as the quantitative measure of the ``compositeness" of the particle, which is given by
\begin{align}
    1-Z_{NR}
    & = \int d\bm{q} |\bra{\bm{q}}B\, \rangle |^2  .
    \nonumber
\end{align}
Multiplying $\bra{\bm{q}}$ to Eq.~\eqref{eq:Schroedinger} from the left, we obtain
\begin{align}
    1-Z_{NR}
    &= 
    \int d\bm{q}
    \frac{|\bra{\bm{q}}V\ket{B}|^2}{[E(q)+B]^2}  .
    \nonumber
\end{align}
Here the matrix element expresses the coupling strength of the bound state to the scattering state with momentum $\bm{q}$. For an $s$-wave bound state, the coupling is independent of the angular variables and is a function of the energy. Then we can write $\bra{\bm{q}}V\ket{B}\equiv G_W(E)$, and obtain the energy integration form
\begin{align}
    1-Z_{NR}
    &=4\pi\sqrt{2\mu^3}
    \int_{0}^{\infty} dE\frac{\sqrt{E}
    |G_W(E)|^2}{(E+B)^2} . 
    \label{eq:1mZexact}
\end{align}
This is an exact expression of the compositeness $1-Z_{NR}$. 

Next we consider to express the integrand of this equation by the scattering amplitude. The formal solution of the Lippmann-Schwinger equation is written as
\begin{align}
    T(E)
    &=V+V\frac{1}{E-H}V  
    . \nonumber
\end{align}
Inserting the complete set of the full Hamiltonian~\eqref{eq:completefull} and taking the matrix element by the initial and final scattering states with energy $E$, we obtain the Low's equation~\cite{Weinberg:1965zz},
\begin{align}
    t(E)
    =&v
    +
    \frac{|G_W(E)|^2}{E+B} 
    +4\pi\sqrt{2\mu^3}\int_0^{\infty} dE^{\prime}
    \frac{\sqrt{E^{\prime}}|t(E^{\prime})|^2}
    {E-E^{\prime}+i\epsilon} ,
    \nonumber
\end{align}
where $v$ ($t$) is the matrix element of the interaction Hamiltonian $V$ ($T$ operator) by the scattering states and we have used $V\ket{\bm{q},\text{full}}=T\ket{\bm{q}}$. The second term of this equation, which comes from the operator $(E-H)^{-1}$ acting on the bound state discrete level, is a part of the integrand of Eq.~\eqref{eq:1mZexact}. Therefore, we can express the compositeness using the scattering amplitude as~\cite{Hyodo:2010uh}
\begin{align}
    1-Z_{NR}
    =&4\pi\sqrt{2\mu^3}
    \int_{0}^{\infty} dE \frac{\sqrt{E}}{E+B}
    \Biggl[
    t(E)-v \nonumber \\
    &
    -
    4\pi\sqrt{2\mu^3}\int_{0}^{\infty} dE^{\prime}
    \frac{\sqrt{E^{\prime}}|t(E^{\prime})|^2}{E-E^{\prime}+i\epsilon}
    \Biggr] .
    \label{eq:compositenessNRexact}
\end{align}
It is important to note that the scattering amplitude $t(E)$ is, in general, complex, but the imaginary part of Eq.~\eqref{eq:compositenessNRexact} vanishes thanks to the optical theorem:
\begin{align}
    \im t(E)
    =&-\tfrac{1}{2} 2\pi\cdot 4\pi\sqrt{2\mu^3}\sqrt{E}|t(E)|^2 
    \nonumber \\
    =&-4\pi^{2} \mu q(E)|t(E)|^2
    \label{eq:opticalnonrel} ,
\end{align}
where $q(E)=\sqrt{2\mu E}$ is the momentum. Therefore, the imaginary part vanishes in the bracket in Eq.~\eqref{eq:compositenessNRexact}, and the overlap $Z_{NR}$ is obtained as a real number. More explicitly, we can write
\begin{align}
    1-Z_{NR}
    =&4\pi\sqrt{2\mu^3}
    \int_{0}^{\infty} dE \frac{\sqrt{E}}{E+B}
    \Biggl[
    \re t(E)-v \nonumber \\
    &
    -
    4\pi\sqrt{2\mu^3}\mathcal{P}\int_{0}^{\infty} dE^{\prime}
    \frac{\sqrt{E^{\prime}}|t(E^{\prime})|^2}{E-E^{\prime}}
    \Biggr] ,
    \nonumber
\end{align}
where $\mathcal{P}$ stands for the principal value integration.

Now we take the limit $B\to 0$. For small $B$, the bracket in Eq.~\eqref{eq:compositenessNRexact} is dominated by the bound state pole term in the $T$ matrix,
\begin{align}
    \text{Integrand in Eq.~\eqref{eq:compositenessNRexact}}
    \approx 
    &
    \frac{\sqrt{E}}{E+B}
    \cdot
    \left(
    \frac{g_W^2}{E+B}
    +\mathcal{O}(B^0)
    \right) , \nonumber \\
    g_W
    \equiv &G_W(E=-B) , \nonumber
\end{align}
because this is the only term with $\mathcal{O}(B^{-1})$ for small $E$. Then the $E$ integration can be done analytically:
\begin{align}
    1-Z_{NR}
    \approx
    &4\pi\sqrt{2\mu^3}g_W^2
    \int_{0}^{\infty} dE \frac{\sqrt{E}}{(E+B)^2} 
    \nonumber \\
    =&
    2\pi^2  
    \sqrt{2\mu^3} \frac{g_W^2}{\sqrt{B}} .
    \label{eq:compositenessNR}
\end{align}
This is the result of Ref.~\cite{Weinberg:1965zz} which connects the compositeness of the bound state with the coupling constant $g_W$ and the binding energy $B$. Since $Z_{NR}\geq 0$, the upper limit of the coupling strength can be obtained as 
\begin{align}
    g_W^2
    &\leq 
    \frac{1}{2\pi^2}  
    \sqrt{\frac{B}{2\mu^3}}  .
    \nonumber
\end{align}
The equality corresponds to $Z_{NR}=0$, namely, the purely composite particle. Note that the upper limit decreases as we decrease the binding energy $B$ and vanishes in the weak binding limit $B\to 0$.

It is important to note that in the expression~\eqref{eq:compositenessNR}, the explicit dependence on the interaction $V$ is lost when $B\to 0$, and all the information of the interaction is represented by the coupling constant of the bound state $g_W$. In this sense, the result is universal and does not depend on the particular choice of the interaction $V$. In contrast, the exact formula~\eqref{eq:compositenessNRexact} can be applied to the bound state with arbitrary binding energy. However, the result depends on the matrix element $v$ and, hence, on the choice of the interaction Hamiltonian $V$. 

From the expression of Eq.~\eqref{eq:1mZexact}, we notice that the exact result of the compositeness can be calculated with the knowledge of the energy dependence of the coupling strength $G_{W}(E)=\bra{\bm{q}}V\ket{B}$. In principle, $G_{W}(E)$ represents the coupling strength of the bound state to the scattering state above the threshold, so it is an off-shell quantity. The coupling ``constant'' $g_{W}$ can be regarded as the leading term in the expansion of $G_{W}(E)$ around $E=-B$. In the present discussion, we assume that the binding energy $B$ and the coupling constant $g_{W}$ are known quantities. In this case, the approximation of $1-Z$ with only known quantities should be in the form of Eq.~\eqref{eq:compositenessNR} and any further contribution should depend on the choice of the interaction $V$ because it is related with the off-shell quantity $G_{W}(E)$. To extend the model-independent formula~\eqref{eq:compositenessNR}, we need to measure the higher-order coefficients of the expansion of $G_{W}(E)$ experimentally.

\subsection{Compositeness in relativistic field theory}
\label{subsec:rel}

In this section, we define the compositeness following the method in Ref.~\cite{PR136.B816}. To define the field renormalization constant, we consider the field theory with a baryon field $\psi(J^{P}=1/2^{+})$, a meson field $\phi(0^{-})$, and a bare field of another baryon $B_0(1/2^{-})$ as
\begin{align}
    \mathcal{L}_0
    =&
    \bar{\psi}(i\Slash{\partial}-M)\psi
    +\tfrac{1}{2}(\partial_{\mu}\phi \partial^{\mu}\phi 
    -m^2\phi^2)
    \nonumber \\
    &+\bar{B}_0 (i\Slash{\partial}-M_{B_0}) B_0 .
    \label{eq:Lagrangian}
\end{align}
We consider the bound state of $\psi$ and $\phi$, which couples with the bare state $B_{0}$. Setting $\mu=Mm/(M+m)$, we obtain the system corresponding to that described by the Hamiltonian $H_{0}$ in the previous section. The interaction Lagrangian, which corresponds to $V$ in the previous section, is introduced as a scalar-type Yukawa form,
\begin{align}
    \mathcal{L}_{\text{int}}
    &=
    g_0\bar{\psi}\phi B_0 + (\text{H.c.}) ,
    \label{eq:int}
\end{align}
where $g_0$ is the bare coupling constant. We then consider that the spectrum of the full theory has a bound state with mass $M_B$ which is related to the binding energy as $M_B=M+m-B$. The definition of the field renormalization constant, which we denote $Z$, is the residue of the full Green's function at $W=M_B$, namely,
\begin{align}
    \Delta(W)
    &=
    \frac{Z}{W-M_B} .
    \label{eq:Zpdef}
\end{align}
For later convenience in the application to the chiral unitary approach, we adopt the positive energy part of the fermion propagator.

Let us calculate this $Z$ constant with the Lagrangian given in Eqs.~\eqref{eq:Lagrangian} and \eqref{eq:int}. The free Green's function for $B_0$ is given by
\begin{align}
    \Delta_0(W)
    &= \frac{1}
    {W-M_{B_0}} .
    \nonumber
\end{align}
The full Green's function $\Delta(W)$ is expressed in terms of $\Delta_0(W)$ by the Dyson equation:
\begin{align}
    \Delta(W)
    &= \Delta_0(W)
    +\Delta_0(W)g_0G(W)g_0\Delta(W)
    \nonumber ,
\end{align}
with the unrenormalized two-body loop function of $\psi$ and $\phi$, $G(W)$. The solution of this equation is 
\begin{align}
    \Delta(W)
    &= \frac{1}
    {[\Delta_0(W)]^{-1}
    -g_0^2G(W)} \nonumber \\
    &= \frac{1}
    {W-M_{B_0}
    -g_0^2G(W)} \nonumber .
\end{align}
The divergence of the loop function $G(W)$ is canceled by the infinite bare mass $M_{B_0}$. Using the renormalized loop function $G(W;a)$ with the parameter $a$ which characterizes the finite part, the full Green's function is given by
\begin{align}
    \Delta(W)
    &= \frac{1}
    {W-g_0^2G(W;a)} 
    \label{eq:Delta} .
\end{align}
The renormalization condition can be obtained by requiring the bound state pole at $W=M_B$.\footnote{This procedure is in principle scale dependent. In the present formulation, the renormalization scale is implicitly set at the baryon mass, $\mu_{s}=M$.} Equating Eq.~\eqref{eq:Delta} with Eq.~\eqref{eq:Zpdef}, we obtain the condition 
\begin{align}
    M_B
    &=
    g_0^2G(M_B;a).
    \nonumber
\end{align}
The field renormalization constant $Z$ can be calculated as
\begin{align}
    Z
    &=
    \lim_{W\to M_B}
    \frac{W-M_B}{W-g_0^2G(W;a)} \nonumber \\
    &= \frac{1}{1-g_0^2G^{\prime}(M_B)}
    \label{eq:Zrel} .
\end{align}

The meson-baryon scattering amplitude is given by $T(W)=g_0\Delta(W)g_0$, so the residue of the amplitude at $W=M_B$ is
\begin{align}
    \lim_{W\to M_B}(W-M_B)T(W)
    &=
    g_0^2Z
    \nonumber .
\end{align}
The residue of the bound-state pole is the physical coupling constant squared, so we obtain the relation
\begin{align}
    g^2
    &=g_0^2Z
    \nonumber 
\end{align}
Using this relation and Eq.~\eqref{eq:Zrel}, we finally write the compositeness $1-Z$ as
\begin{align}
    1-Z
    &=
    -g^2G^{\prime}(M_B) \label{eq:compositenessR} .
\end{align}
Note that the right-hand side is expressed only by the physical (renormalized) quantities, and no approximation has been applied to the evaluation of the mass of the bound state. Thus, this is an exact expression of the compositeness of the bound state in the relativistic field theory with the interaction Lagrangian~\eqref{eq:int}, which is expressed in terms of the coupling strength $g$ of the physical bound state to the scattering state and the derivative of the loop function at the energy of the bound state. The expression for the scalar meson bound state in the meson-meson scattering is derived in Appendix~\ref{sec:scalar}. It should be noted that the definition of the compositeness~\eqref{eq:compositenessR} depends on the adopted interaction Lagrangian $\mathcal{L}_{\text{int}}$. With a different interaction, the function form of Eq.~\eqref{eq:compositenessR} will be modified accordingly, and, hence, the resulting expression for $1-Z$ will be changed.

The possible region of the bound state mass is $M\leq M_B\leq M+m$, where $G^{\prime}(M_B)$ is real and negative, and the coupling square is always positive for the physical bound state. Therefore, we find that the compositeness is always positive,
\begin{align}
    1-Z
    &\geq 0 .\label{eq:lowerlimit}
\end{align}
In other words, the field renormalization constant is smaller than unity.

The field renromalization constant is well defined for resonances. Using the same argument, we find
\begin{align}
    1-Z
    &=
    -g^2G_{II}^{\prime}(z_{R}) \label{eq:compositenessresonance} ,
\end{align}
where $g^{2}$ is the residue of the pole in the complex energy plane of the scattering amplitude. This is a natural extension of Eq.~\eqref{eq:compositenessR}. However, it is not straightforward to interpret Eq.~\eqref{eq:compositenessresonance} as the compositeness, because the condition~\eqref{eq:lowerlimit} is not always guaranteed and $1-Z$ may become a complex number. Nevertheless, Eq.~\eqref{eq:compositenessresonance} may be applied to  a narrow width resonance, because it coincide with Eq.~\eqref{eq:compositenessR} in the zero width limit.

\section{Compositeness of dynamically generated states}\label{sec:numerical}

\subsection{Application to chiral unitary approach}\label{subsec:application}

Here we analyze the compositeness of the bound state in a specific model described in section~\ref{sec:ChU}. In the present section, we have derived three definitions of the compositeness: 
\begin{itemize}
\item[(i)] expression in nonrelativistic formalism, Eq.~\eqref{eq:compositenessNRexact};
\item[(ii)] formula for small $B$, Eq.~\eqref{eq:compositenessNR}; 
\item[(iii)] expression in relativistic formalism, Eq.~\eqref{eq:compositenessR}.
\end{itemize}
For the application to the chiral unitary approach with the energy-dependent interaction, it is fully legitimate to use Eq.~\eqref{eq:compositenessR}. In the nonrelativistic formalism, we should choose the framework which can deal with the energy-dependent interaction. In this respect, the exact result Eq.~\eqref{eq:compositenessNRexact} cannot be used because the completeness relation~\eqref{eq:completefull} is not always valid. However, the model-independent result for small $B$~\eqref{eq:compositenessNR} may be appropriate because it can be derived without using the completeness of the full Hamiltonian~\cite{Weinberg:1965zz}. Thus, in the following, we examine the expression of the compositeness in the relativistic field theory in Eq.~\eqref{eq:compositenessR} and the model-independent result for small binding energy, Eq.~\eqref{eq:compositenessNR}.

Substituting Eqs.~\eqref{eq:Eindepcoupling} and \eqref{eq:WTcoupling} into the exact result in the field theory~\eqref{eq:compositenessR}, we obtain
\begin{align}
    &1-Z 
    =
    \begin{cases}
    1
    & \text{constant interaction}, \\
    \dfrac{1}
    {
    1+\frac{G(M_B;a)}{(M_B-M)G^{\prime}(M_B)}
    }
    & \text{WT interaction}.
    \end{cases}
    \label{eq:1mZR}
\end{align}
It is striking that the constant interaction always produces a purely composite bound state. This corresponds to the equivalence of the four-Fermi theory and Yukawa theory found in Ref.~\cite{PR136.B816}. On the other hand, the compositeness of the bound state generated by the WT interaction depends on the subtraction constant. This result agrees with the fact of the appearance of the pole term in the effective interaction for the WT interaction~\eqref{eq:poleterm}, which introduces the source of the elementarity. For a general interaction $V$, using Eq.~\eqref{eq:generalg2}, we obtain
\begin{align}
    1-Z 
    =& 
    \frac{1}
    {
    1-\frac{V^{\prime}(M_B)G(M_B)}{V(M_B)G^{\prime}(M_B)}
    }
    \label{eq:1mZgeneral} ,
\end{align}
where $V^{\prime}(M_{B})=\partial V(W)/\partial W|_{W=M_{B}}$. In this expression, $G$ and $V$ are determined under a fixed renormalization scheme. It is easy to see that Eq.~\eqref{eq:1mZR} follows from the general expression~\eqref{eq:1mZgeneral} with the interaction kernel~\eqref{eq:WTinteraction} and \eqref{eq:Eindepinteraction} in the phenomenological renormalization scheme. In addition, the energy dependence of the interaction kernel is related to the deviation of the compositeness from unity.

Let us recall the definition of the bound state that $M_B-M\geq 0$ and the behavior of the loop function below the threshold indicates $G^{\prime}(M_B)< 0$. Relation~\eqref{eq:WTstrength} implies $G(M_B;a)\leq 0$, so altogether we obtain
\begin{align}
    \dfrac{G(M_B;a)}{(M_B-M)G^{\prime}(M_B)}
    &\geq 
    0 , \nonumber
\end{align}
and, hence, with Eq.~\eqref{eq:lowerlimit}, for the WT interaction, we obtain
\begin{align}
    0 \leq 
    1-Z
    \leq 1 .
    \nonumber
\end{align}
This result indicates that the compositeness defined in Eq.~\eqref{eq:1mZR} is normalized within the range from zero to one. This is a very useful aspect to discuss the structure of the bound state in a quantitative manner. 

For the WT interaction, a purely composite bound state ($Z=0$) is achieved when
\begin{align}
    G^{\prime}(M_B)
    &=
    -\infty \quad
    \text{or} \quad
    G(M_B;a)
    =
    0
    \label{eq:compositenesscondition} .
\end{align}
Because the derivative of the loop function diverges at threshold, the first condition can be met if the bound state is produced exactly at the threshold, irrespective of the value of the subtraction constant. For $M_B\neq M+m$, a purely composite bound state would be produced only if $G(M_B;a)$ is zero, which indicates $C\to -\infty$ by Eq.~\eqref{eq:WTstrength}. Thus, to realize $Z=0$, we should have 
\begin{align}
    M_B
    &=
    M+m \quad
    \text{or} \quad
    C
    \to
    -\infty
    \label{eq:compositenesscondition2} .
\end{align}
Note that the condition~\eqref{eq:compositenesscondition2} is only satisfied in the limiting cases so it implies that the physical bound state (finite binding energy and finite interaction strength) cannot be purely composite for the WT interaction. It is instructive to recall that the natural renormalization condition developed in Ref.~\cite{Hyodo:2008xr} requires $G(M;a_{\text{natural}})=0$. Interestingly, this condition was introduced to realize the hadronic molecule state in the amplitude, and agrees with Eq.~\eqref{eq:compositenesscondition} in the chiral limit ($m\to 0$). In the next section we numerically study the compositeness of the bound state generated in the natural renormalization scheme.

Next we consider the small binding limit. Taking into account the convention of the scattering amplitude as summarized in Appendix~\ref{sec:convention}, we write the coupling constant $g_W$ in Eq.~\eqref{eq:compositenessNR} in terms of the coupling strength in the chiral unitary approach as
\begin{align}
    g_W^2
    &= 
    \frac{M}{16\pi^3\mu M_B}\frac{\bar{q}(M_B)}{q(-B)}
    g^2 .
    \nonumber
\end{align}
Note that both the momentum factors $q(-B)$ and $\bar{q}(M_B)$ are pure imaginary. Substituting this expression into Eq.~\eqref{eq:compositenessNR}, we obtain~\cite{Hyodo:2010uh}
\begin{align}
    1-Z_{NR}
    &= 
    \frac{M|\lambda^{1/2}(M_B^2,M^2,m^2)|}{16\pi M_B^2(M+m-M_B)}
    g^2
    , \label{eq:1mZNR}
\end{align}
for $B\to 0$.
This form, in comparison with the relativistic result~\eqref{eq:compositenessR}, leads to the relation
\begin{align}
    1-Z
    & \approx
    \left(
    \frac{-16\pi M_B^2(M+m-M_B)G^{\prime}(M_B)}
    {M|\lambda^{1/2}(M_B^2,M^2,m^2)|}
    \right)
    (1-Z_{NR}) \nonumber \\
    &\equiv 
    A(M_B)(1-Z_{NR}) \label{eq:Adef}
\end{align}
for the small binding energy. Note that the factor $A(M_B)$ depends only on the mass of the bound state. In the small binding limit, the definition of the field renormalization constant $Z$ should be consistent with that of $Z_{NR}$, so we expect
\begin{align}
    \lim_{M_B\to M+m}A(M_B) 
    &=1 .
    \nonumber
\end{align}
We will check the validity of this relation in the numerical calculation.

Finally we briefly mention the compositeness for virtual states and resonances. The present formulation in the nonrelativistic quantum mechanics is not directly applicable to resonances which are not included in the complete set in Eq.~\eqref{eq:complete} and in Eq.~\eqref{eq:completefull}.\footnote{The resonance state can be included in the extended completeness relation using the complex scaling method~\cite{PTP98.1275}. This is one possibility to define the compositeness of resonances.} However, the residue of the resonance propagator can be evaluated without trouble in the relativistic field theory. This provides a natural extension of the compositeness of bound states to virtual states and resonances. Equation~\eqref{eq:compositenessresonance} leads to 
\begin{align}
    1-Z
    &=\begin{cases}
    1
    & \text{constant interaction}, \\
    \dfrac{1}
    {
    1+\frac{G_{II}(z_R;a)}{G^{\prime}_{II}(z_R)(z_R-M)}
    }
    & \text{WT interaction}.
    \end{cases} \label{eq:compositenessRresonance} 
\end{align}
For a virtual state ($\im z_{R}=0$), the expression provides a real value for the compositeness. The virtual state generated by the contact interaction always gives $Z=0$, while the WT interaction can lead to $Z<0$, depending on the subtraction constant. This is because the virtual state cannot be included in the orthonormal basis. In the case of a resonance, the field renormalization constant is given by a complex number. Thus, the interpretation of the field renormalization constant as the compositeness is not as straightforward as in the case of the bound state. However, when the width is small, the imaginary part of the renormalization constant is small and we can read off the compositeness from its real part or from its absolute value.

\subsection{Numerical analysis}\label{subsec:numerical}

In this section we present numerical results. To fix the energy scale of the system, we choose $m=496$ MeV and $M=939$ MeV unless otherwise stated, bearing the $\bar{K}N$ bound state~\cite{Hyodo:2007jq} in mind.

We first calculate the compositeness of the bound states obtained in the natural renormalization condition to examine the nature of the bound state in this scheme. The natural renormalization scheme is briefly explained in Appendix~\ref{sec:generalcase}. As shown in Eq.~\eqref{eq:1mZR}, the constant interaction always produces a purely composite bound state, so we investigate the case of the energy-dependent WT interaction. We set $a=a_{\text{natural}}$ and vary the binding energy $B=M+m-M_B$ with the coupling strength $C$ being determined according to Eq.~\eqref{eq:WTstrength}. 

\begin{figure}[tbp]
\includegraphics[width=7cm,clip]{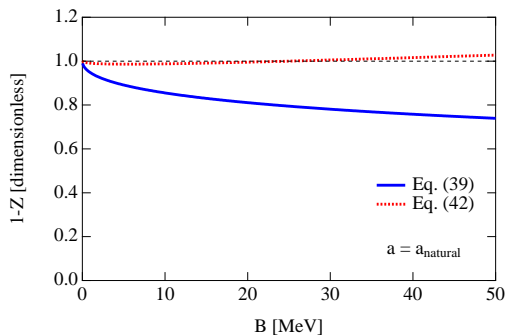}
\caption{\label{fig:CompositenessB} (Color online) The compositeness of the bound state as a function of the binding energy $B=M+m-M_B$ with $a=a_{\text{natural}}$, $m=496$ MeV, and $M=939$ MeV. The solid line denotes expression~\eqref{eq:1mZR}, while the dotted line represents the approximate form~\eqref{eq:1mZNR} valid for the small binding energy.}
\end{figure}

In Fig.~\ref{fig:CompositenessB}, we show the compositeness $1-Z$ given in Eq.~\eqref{eq:1mZR} by the solid line as a function of the binding energy $B$. The result indicates that the generated bound state in the natural renormalization scheme is dominated by the composite structure. When the binding energy is decreased, the compositeness of the bound state becomes large and eventually approaches unity for $B\to 0$, in accordance with Eq.~\eqref{eq:compositenesscondition2}. The deviation from the purely composite state becomes larger as the binding energy is increased. In practice, to produce the deeply bound state in the natural scheme, the coupling strength $C$ must be very large, which is unlikely when this strength follows the constraint from chiral symmetry~\cite{Hyodo:2006yk,Hyodo:2006kg}.

\begin{figure}[tbp]
\includegraphics[width=7cm,clip]{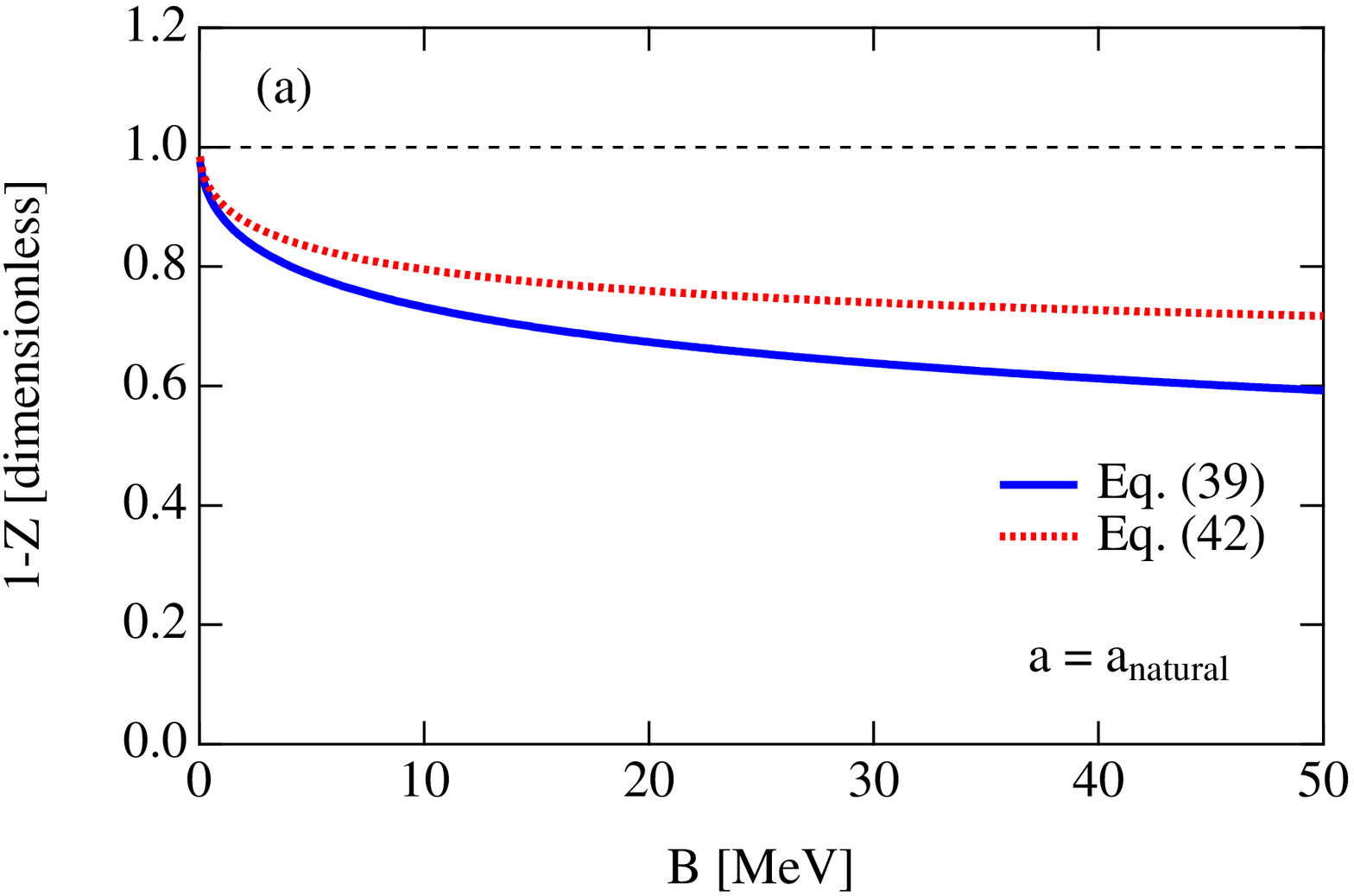}
\includegraphics[width=7cm,clip]{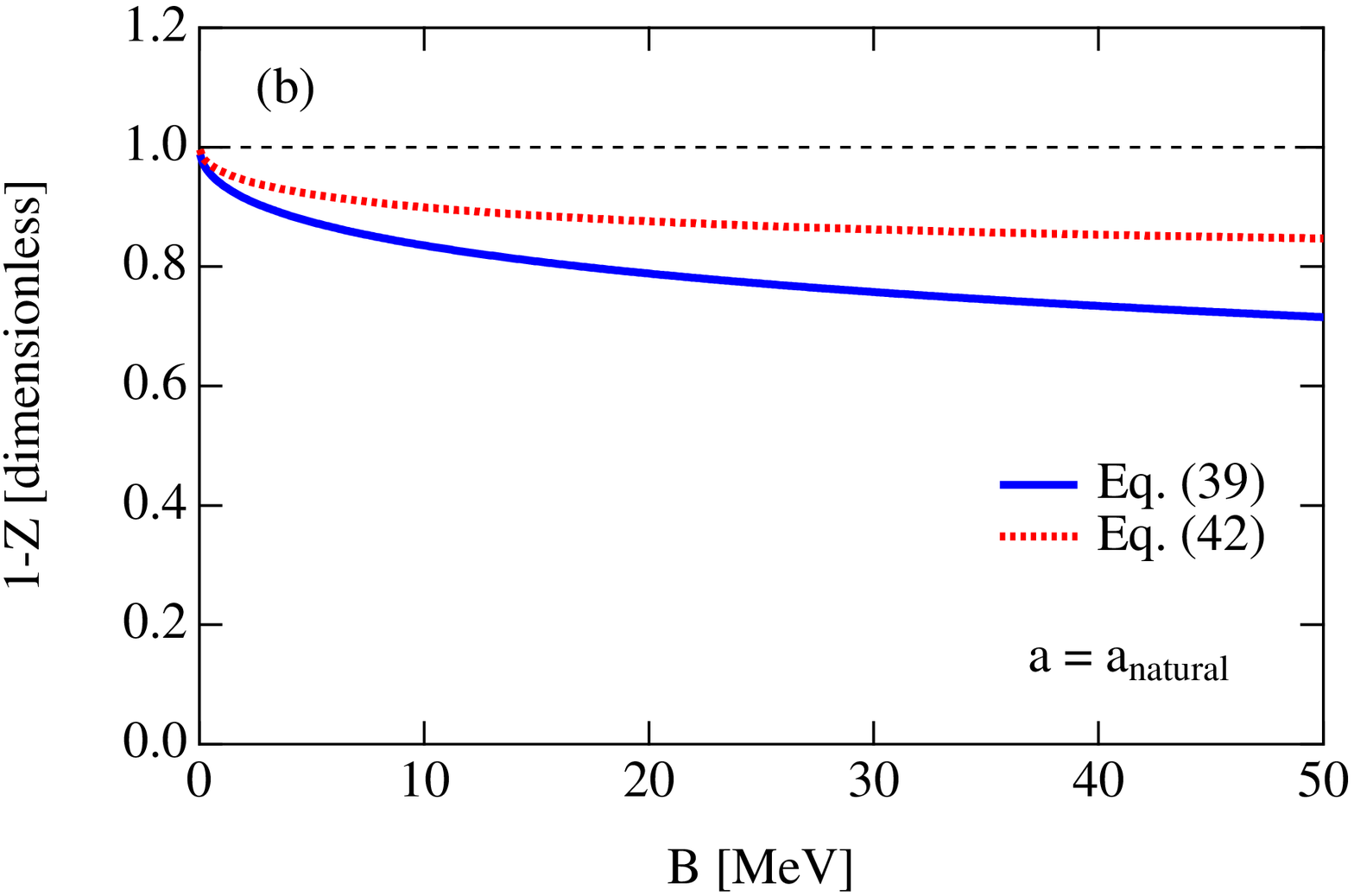}
\includegraphics[width=7cm,clip]{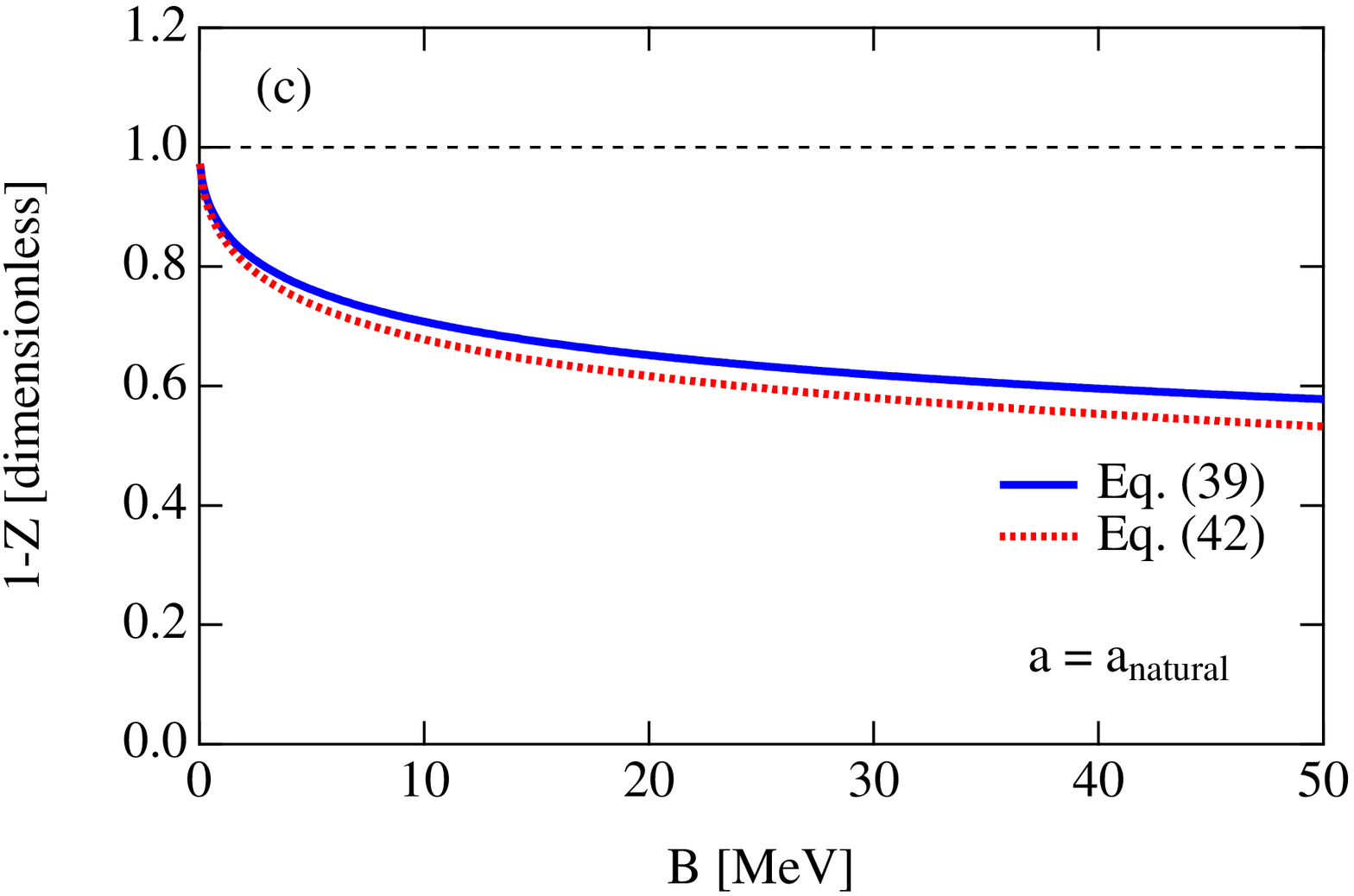}
\caption{\label{fig:CompositenessB_other} (Color online) Same with Fig.~\ref{fig:CompositenessB} with $m=138$ MeV and $M=939$ MeV [$\pi N$ system, panel (a)], $m=496$ MeV and $M=2286$ MeV [$\bar{K} \Lambda_{c}$ system, panel (b)], and $m=138$ MeV and $M=2286$ MeV [$\pi \Lambda_{c}$ system, panel (c)].}
\end{figure}

For comparison, we plot the result of the model-independent form~\eqref{eq:1mZNR} with the dotted line, which is valid for the small binding energy. We notice that the result of the small binding approximation deviates from the result of Eq.~\eqref{eq:1mZR}, even for relatively small binding energy, although the compositeness becomes unity in the $B\to0$ limit. In addition, for the binding energy $B>25$ MeV, the compositeness exceeds unity, which means $Z<0$. In comparison with the exact result, we conclude that the result $Z<0$ is caused by the breakdown of the small binding approximation. In Fig.~\ref{fig:CompositenessB_other}, we also present the results with other combination of hadron masses, $m=138$ MeV and $M=939$ MeV [$\pi N$ system, panel (a)], $m=496$ MeV and $M=2286$ MeV [$\bar{K} \Lambda_{c}$ system, panel (b)], and $m=138$ MeV and $M=2286$ MeV [$\bar{K} \Lambda_{c}$ system, panel (c)]. It is seen that the agreement of the exact formula with the model-independent form depends on the combination of hadron masses.

\begin{figure}[tbp]
\includegraphics[width=7cm,clip]{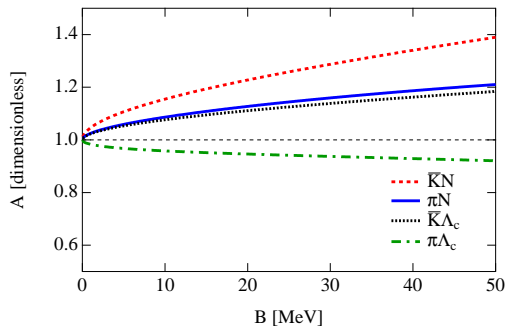}
\caption{\label{fig:Afactor} (Color online) The factor $A(M_B)$ defined in Eq.~\eqref{eq:Adef} which shows the accuracy of the approximation. The dashed line represents the result with $m=496$ MeV and $M=939$ MeV ($\bar{K}N$ system), the solid line with $m=138$ MeV and $M=939$ MeV ($\pi N$ system), the dotted line with $m=496$ MeV and $M=2286$ MeV ($\bar{K} \Lambda_c$ system), and the dash-dotted line with $m=138$ MeV and $M=2286$ MeV ($\pi \Lambda_c$ system).}
\end{figure}

To examine the validity of the small binding approximation, we show in Fig.~\ref{fig:Afactor} the factor $A(M_B)$ in Eq.~\eqref{eq:Adef} which characterizes the difference of Eqs.~\eqref{eq:1mZR} and~\eqref{eq:1mZNR}. To study the dependence on the masses of the system, we analyze the $\bar{K}N$ system, the $\pi N$ system, the $\bar{K} \Lambda_c$ system, and the $\pi\Lambda_{c}$ system. In the $\bar{K}N$, $\pi N$ and $\bar{K}\Lambda_{c}$ cases, the factor $A(M_B)$ enhances the compositeness from the exact value, while it is reduced in the $\pi\Lambda_{c}$ case. In all cases, it becomes unity in the limit $B\to 0$. This means that the field renormalization constant $Z$ in Sec.~\ref{subsec:rel} and $Z_{NR}$ in Sec.~\ref{subsec:nonrel} are equivalent at the zero binding energy:
\begin{align}
    (Z-Z_{NR})\bigl|_{B\to 0}
    =
    0
    \nonumber .
\end{align}
The agreement is, however, only guaranteed in the $B\to 0$ limit, because we include only the leading term in the pole dominance approximation to derive $Z_{NR}$, as discussed in the end of Sec.~\ref{subsec:nonrel}. The deviation of the results for $B\neq 0$ is attributed to the lack of knowledge of the off-shell dependence of the coupling strength $G_{W}$.

\begin{figure}[tbp]
\includegraphics[width=7cm,clip]{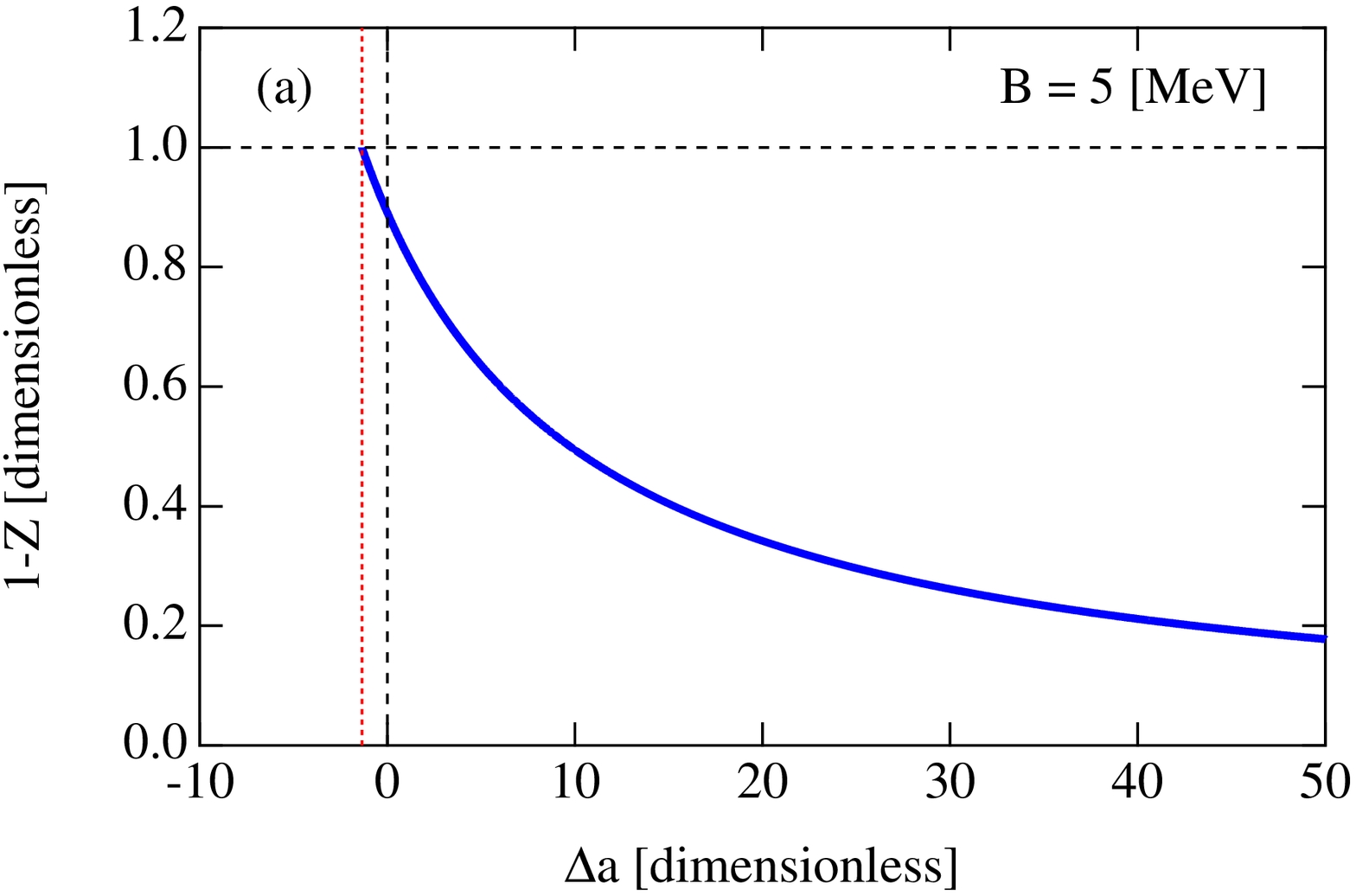}
\includegraphics[width=7cm,clip]{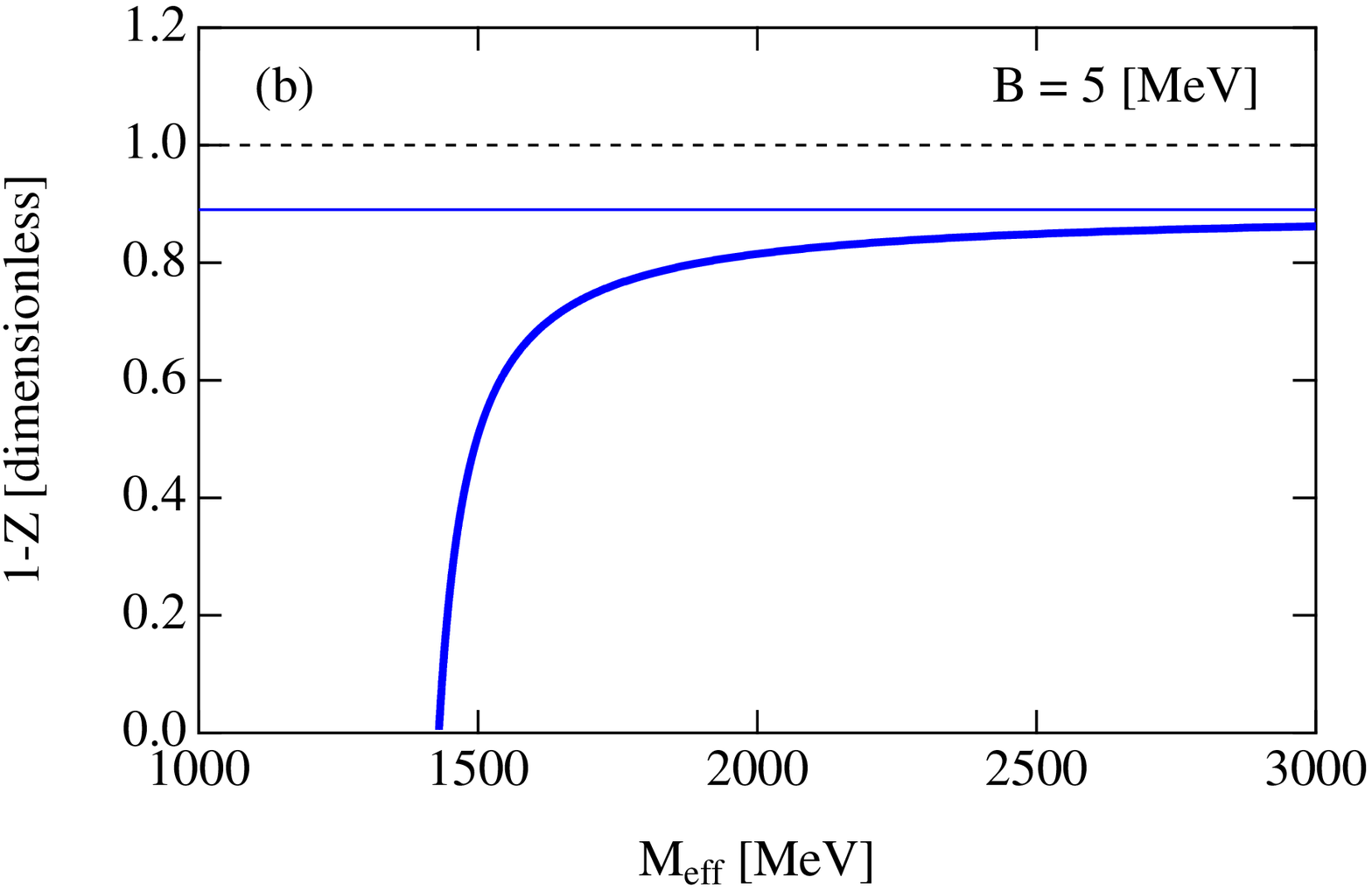}
\caption{\label{fig:Compositenessa} (Color online) The compositeness of the bound state as a function of the deviation of the subtraction constant from the natural value $\Delta a$ (a) and the effective pole mass $M_{\text{eff}}$ (b) with $B=10$ MeV, $m=496$ MeV and $M=939$ MeV. The vertical dotted line represents the composite subtraction constant $a_{\text{comp}}$, while the vertical dashed line in the panel (a) corresponds to the natural value of the subtraction constant $a_{\text{natural}}$. The horizontal solid line in the panel (b) indicates the compositeness at $a=a_{\text{natural}}$ ($M_{\text{eff}}\to \infty$).}
\end{figure}

Next we connect the discussion of the natural renormalization scheme in Ref.~\cite{Hyodo:2008xr} to the compositeness. From now on we use Eq.~\eqref{eq:1mZR} to calculate the compositeness. We fix the binding energy at $B=5$ MeV and vary the subtraction constant from the natural value. Again, the coupling strength $C$ is adjusted using Eq.~\eqref{eq:WTstrength}.  In Fig.~\ref{fig:Compositenessa}(a), we plot the compositeness as a function of the deviation of the subtraction constant from the natural value:
\begin{align}
    \Delta a
    =
    a-a_{\text{natural}}
    \nonumber .
\end{align}
The result indicates that the compositeness of the bound state decreases as the deviation from the natural value gets larger, as discussed in Ref.~\cite{Hyodo:2008xr}. The vertical dotted line corresponds to $G(M_B;a)=0$, where the coupling strength $C$ diverges.

To appreciate the deviation of the subtraction constant in the physical unit, using Eq.~\eqref{eq:Meff}, we translate $\Delta a$ into the effective mass of the pole term in the interaction, $M_{\text{eff}}$. We have set the problem such that the bound state with $B=5$ MeV is generated by the WT term with $a_{\text{natural}}$. To generate the bound state with the same binding energy with a different subtraction constant $a=a_{\text{natural}}+\Delta a$, we take the different coupling strength $C^{\prime}=(M_{B}-M)G(M_{B};a_{\text{natural}}+\Delta a)$. We can express the difference of the subtraction constants in terms of the effective pole term as in Eq.~\eqref{eq:poleterm} by shifting the subtraction constant to the natural value. We calculate the effective pole mass $M_{\text{eff}}$ in Eq.~\eqref{eq:Meff}. The compositeness is displayed as a function of $M_{\text{eff}}$ in Fig.~\ref{fig:Compositenessa}(b). It is clear from Fig.~\ref{fig:Compositenessa}(b) that the bound state becomes purely elementary particle for $M_{\text{eff}}\to M_B$, which corresponds to $\Delta a\to \infty$ by Eq.~\eqref{eq:limit2}. The compositeness of the bound state is large when $M_{\text{eff}}$ is located away from the bound-state energy region, and in the limit $M_{\text{eff}}\to \infty$, by Eq.~\eqref{eq:limit2} we obtain the value with $a=a_{\text{natural}}$ (horizontal solid line). This behavior is fully compatible with the analysis in Ref.~\cite{Hyodo:2008xr}.

Having the compositeness $1-Z$ defined, we revisit the meaning of the natural renormalization scheme. We have found that the physical bound state generated by the WT interaction cannot be purely composite. Thus, even in the natural renormalization scheme, the bound state should contain some elementary component. Because the purely composite state is realized only for the limiting cases, we cannot establish a ``composite renormalization condition'' in which the elementary component is completely excluded. It is possible to calculate the compositeness for the bound state when a phenomenological model is available, but the notion cannot be applied to the scattering system without experimental data, because the knowledge of $M_{B}$ is necessary to calculate $Z$. The numerical results in Figs.~\ref{fig:CompositenessB} and \ref{fig:CompositenessB_other} show that the natural renormalization scheme provides an almost composite bound state for a small binding energy, so we can regard the natural renormalization scheme as a practical way to generate the hadronic molecular states.

It is instructive to recall the phenomenological discussion of Ref.~\cite{Hyodo:2008xr} where the $\Lambda(1405)$ and $N(1535)$ resonances were analyzed. Through the comparison of the phenomenological and natural schemes, it was found that the effective pole mass for $\Lambda(1405)$ appears at 7.9 GeV, while that for $N(1535)$ sits at 1.7 GeV. The result in Fig.~\ref{fig:Compositenessa}(b) implies that more than 80 \% of $\Lambda(1405)$, described as $\bar{K}N$ quasibound state with 5-MeV binding, is interpreted as a meson-baryon composite structure. For more quantitative discussion on the structure of $N(1535)$, we need to extend the present framework to the resonance state in the multi-channel scattering. This is briefly discussed in Appendix~\ref{subsec:Zcoupledbound}.

\section{Summary}\label{sec:summary}

We have discussed the compositeness of a meson-baryon bound state in the single-channel meson-baryon scattering. Based on this model space, we decompose the bound-state wave function into the meson-baryon composite system and the elementary contribution expressed by the bare field. Assuming that the mass of the bound state and its coupling to the scattering state are known, we determine the compositeness of the bound state. The notion of the compositeness is used to investigate the structure of the bound state obtained in the chiral unitary approach.

We first study the properties of the bound states and resonances in the chiral unitary approach. Analyzing the single-channel model with the energy-independent constant interaction and the energy-dependent chiral interaction, we show that the residue of the bound-state pole, which expresses the squared coupling constant, is always positive irrespective to the energy dependence of the interaction. Using the pole condition, we find that a pole with a finite imaginary part (a resonance or a virtual state with a finite width) can be generated when the interaction has energy dependence, in the framework of the chiral unitary approach.

The normalized measure of the compositeness of the bound state is defined in a general framework, through the renormalization constant of the bare field. The compositeness is expressed in terms of the mass of the bound state and the coupling constant to the scattering state. We find the expression of the compositeness for arbitrary binding energy in the nonrelativistic quantum mechanics and in the relativistic field theory, from which the model-independent formula of Weinberg~\cite{Weinberg:1965zz} is derived in the weak binding limit. It is, however, shown that the compositeness of the bound state, except for the weak binding limit, depends on the choice of the interaction, unless the off-shellness of the coupling strength of the bound state to the scattering states is under control.

Applying this formulation to the chiral unitary approach, we find that the constant interaction always generates a purely composite bound state. In contrast, when the energy-dependent chiral interaction is used, the bound state is realized as a mixture of the elementary and composite components, and a purely composite bound state is possible in the limit of vanishing binding energy $B\to 0$ or diverging interaction strength $C\to -\infty$.

We show that the bound state in the natural renormalization scheme~\cite{Hyodo:2008xr} is almost composite particle. In Ref.~\cite{Hyodo:2008xr}, the pole term in the effective interaction was interpreted as the source of elementarity. This picture is confirmed from the viewpoint of the compositeness of the bound state, and the position of the effective pole is now related to the compositeness quantitatively. 

Given that the most of the interesting hadrons are resonances, it is of great importance to generalize the notion of the ``compositeness" to resonances. Although a naive extension of the present approach \eqref{eq:compositenessRresonance} results in the complex value for the compositeness, a quantitative definition of the compositeness of resonances will be an important step to understand the structure of hadron resonances.

\begin{acknowledgments}
The authors are grateful to Koichi Yazaki, Hideko Nagahiro, Kanabu Nawa, and Sho Ozaki for fruitful discussions and suggestions. T.H. thanks Koji Harada and particle theory group at Kyushu University for useful discussions.
T.H. is grateful the support from the Global Center of Excellence Program by MEXT, 
Japan through the Nanoscience and Quantum Physics Project of the Tokyo Institute 
of Technology. 
This work was partly supported by the Grant-in-Aid for Scientific Research from 
MEXT and JSPS (Grants No. 22740161,  No. 22105507, 
and No. 21840026) 
and the Grant-in-Aid for the Global COE Program ``The Next Generation of Physics, Spun 
from Universality and Emergence'' from MEXT, Japan.
This work was done in part under the Yukawa International Program for Quark-hadron Sciences (YIPQS).
A.H. is supported in part by the Grant-in-Aid for Scientific
Research on Priority Areas entitled ``Elucidation of New
Hadrons with a Variety of Flavors'' (Grant No. E01: 21105006).
\end{acknowledgments}

\appendix

\section{CONVENTION OF THE AMPLITUDE}\label{sec:convention}

Here we summarize the convention of the scattering amplitude and $T$ matrix in this paper (see also Ref.~\cite{Hyodo:2011ur}). The scattering amplitude in the nonrelativistic quantum mechanics $f(E,\theta)$ with the energy $E$ and the scattering angle $\theta$ is related to the differential cross section as
\begin{align}
    |f(E,\theta)|^2 
    =
    \frac{d\sigma(E)}{d\Omega} .
    \nonumber
\end{align}
For the $s$-wave scattering without angular dependence, the optical theorem can be written as
\begin{align}
    \im f(E)
    =
    q |f(E)|^2 ,
    \nonumber
\end{align}
where $q$ is the momentum $q=\sqrt{2\mu E}$. The $T$-matrix element in the nonrelativistic quantum mechanics $t$ is defined as
\begin{align}
    f(E)
    =
    -4\pi^2 \mu t(E) ,
    \nonumber
\end{align}
such that the optical theorem takes the form of Eq.~\eqref{eq:opticalnonrel}.

The $T$ matrix in the chiral unitary approach $T$ satisfies the optical theorem as
\begin{align}
    \im T(W)
    =
    -\frac{M \bar{q} }{4\pi W} |T(W)|^2 ,
    \nonumber
\end{align}
where the three-momentum function $\bar{q}$ is defined in Eq.~\eqref{eq:barq} and $W=E+M+m$. Therefore, to have the consistent optical theorem, the relation between $T$ and $t$ should be as follows:
\begin{align}
    T(W)
    =
    -\frac{4\pi \sqrt{s}}{M }\frac{q}{\bar{q}} f(E)
    =\frac{16\pi^3\mu \sqrt{s}}{M }\frac{q}{\bar{q}}t(E) .
    \nonumber
\end{align}

\section{NATURAL RENORMALIZATION SCHEME AND EFFECTIVE POLE TERMS FOR GENERAL INTERACTION}\label{sec:generalcase}

In Ref.~\cite{Hyodo:2008xr}, the natural renormalization scheme was introduced to study the origin of resonances, 
based on the property of the real part of the loop function and the chiral low energy theorem. The subtraction constant $a$ in Eq.~\eqref{eq:Gfn} determines the finite part of the loop function. To exclude the possible seed of resonances in the loop function, the subtraction constant should be chosen to satisfy the condition
\begin{align}
    G(\mu_{m};a_{\text{natural}}) 
    &=0,\quad
    \mu_{m}=M .
    \label{eq:natural}
\end{align}
This subtraction constant $a_{\text{natural}}$ is called the natural subtraction constant. As shown in Eqs.~\eqref{eq:poleterm} and \eqref{eq:Meff}, the variation of the subtraction constant (with the same interaction kernel) is equivalent to the introduction of the pole term in the interaction kernel (with the same subtraction constant). Thus, if the experimental data requires the subtraction constant which is a very different value from the natural one, the origin of the generated resonance should be attributed to be the CDD pole contribution. However, if the experimental data are well reproduced by  the natural renormalization constant, then the resonance in the amplitude is regarded as a dynamically generated state.

Let us consider the relation between the deviation of the subtraction constant $\Delta a=a-a_{\text{natural}}$ and the effective mass $M_{\text{eff}}$. From Eq.~\eqref{eq:Meff}, we find
\begin{align}
    M_{\text{eff}}\to \infty
    \quad &\Leftrightarrow \quad
    \Delta a=0 \label{eq:limit1}  .
\end{align}
This can be understood to mean that the natural value of the subtraction constant does not require the pole term in the interaction kernel.  If there is no bound state, we obtain $M_{\text{eff}}=M$ from Eq.~\eqref{eq:Meff} in the limit of $\Delta a\to \infty$. If we have a bound state at $W=M_{B}$, however, this does not hold as shown below. Eliminating the coupling strength $C$ in Eq.~\eqref{eq:Meff} by the bound state condition~\eqref{eq:WTcond}, we find
\begin{align}
   M_{\text{eff}}
    =&
    M+\frac{(4\pi)^2}{2M\Delta a} 
    (M_{B}-M)
    G(M_{B};a_{\text{natural}}+\Delta a)
    \nonumber ,
\end{align}
where $G(M_{B};a_{\text{natural}}+\Delta a)$ becomes infinite as $\Delta a\to \infty$. Noting that the subtraction constant is included linearly in $G(W;a)$, we can write
\begin{align}
    G(M_{B};a_{\text{natural}}+\Delta a)
    = \frac{2M\Delta a}{(4\pi)^2}+G(M_{B};a_{\text{natural}}) .
    \nonumber
\end{align}
Then we obtain
\begin{align}
    M_{\text{eff}}
    =& M_B+\frac{(4\pi)^2}{2M\Delta a}
    (M_{B}-M)G(M_B;a_{\text{natural}}) .
    \nonumber
\end{align}
Because $G(M_B;a_{\text{natural}})$ is finite, we obtain
\begin{align}
    M_{\text{eff}}= M_B
    \quad &\Leftrightarrow \quad
    \Delta a\to \infty .
    \label{eq:limit2}
\end{align}

In Sec. II, we have discussed the interaction $V$ which linearly depends on the meson energy, with single subtraction of the dispersion integral. In general, the higher-order contributions to $V$ in chiral perturbation theory will induce terms with higher powers of meson energy in the interaction kernel, and more than single subtraction can be performed. Here we consider the natural renormalization condition and the effective pole(s) in general cases.

We write the interaction $V$ of $N$ powers of meson energy and the loop function $G$ with $M$ times subtractions as
\begin{align}
    V(W)
    =&\sum_{n=0}^{N} c^{(n)}W^n ,\nonumber \\
    G(W;\{a^{(m)}\})
    =&\sum_{m=0}^{M} a^{(m)}W^m
    +G^{\text{finite}}(W) ,\nonumber \\
    G^{\text{finite}}(W)
    \equiv & G(W;a)-\frac{2Ma}{(4\pi)^2} .\nonumber 
\end{align}
For instance, we can choose $N=0$, $c^{(0)}=Cm$ for the energy-independent interaction, $N=1$, $c^{(0)}=-CM$, $c^{(1)}=C$ for the WT interaction, and $M=0$, $a^{(0)}=-2Ma/(4\pi)^{2}$ for the single subtraction of the loop function.

We first consider the natural renormalization scheme in this case. Because we have $M+1$ subtraction constants, $M+1$ relations should be determined in the natural scheme. In addition to  condition~\eqref{eq:natural},
\begin{align}
    T(\mu_m)
    =&V(\mu_m)
    \nonumber ,
\end{align}
we require the matching of the derivatives
\begin{align}
    \left.
    \frac{\partial^{m} T(W)}{\partial W^{m}}
    \right|_{W=\mu_{m}}
    =&
    \left.
    \frac{\partial^{m} V(W)}{\partial W^{m}}
    \right|_{W=\mu_{m}}
    \quad m= 1,\dots , M
    \nonumber ,
\end{align}
where $\mu_{m}$ is the matching scale which is taken to be the mass of the baryon in Ref.~\cite{Hyodo:2008xr}. These conditions determine the natural subtraction constants $\{a^{(m)}_{\text{natural}}\}$. Then the effective interaction is given by
\begin{align}
    V_{\text{natural}}(W;\{a^{(m)}\})
    =&
    \frac{1}{(\sum_{n=0}^{N} c^{(n)}W^n)^{-1}
    -\sum_{m=0}^{M} a^{(m)}W^m}
    \nonumber \\
    =& \frac{\sum_{n=0}^{N} c^{(n)}W^n}
    {c^{(N)}a^{(M)}
    \prod_{l=1}^{M+N}
    (W-\alpha_{l})}
    \nonumber .
\end{align}
Thus, the effective interaction has $M+N$ poles. The energy-independent interaction is $N=M=0$, so no pole appears, and the WT interaction is $N=1$ and $M=0$ so we have one pole in Eq.~\eqref{eq:Meff}. Because $\{\alpha_{l}\}$ are the roots of the $M+N$th algebraic equation, they can become complex numbers.

\section{POLE TRAJECTORY IN THE SECOND RIEMANN SHEET}
\label{sec:pole}

Here we consider the trajectory of pole singularity in the scattering amplitude when the coupling strength is varied.

We first show that the pole trajectory is continuous as a function of the coupling strength in the second Riemann sheet. Defining\footnote{In this Appendix we fix the subtraction constant $a$ as a natural value and do not consider its variation.} 
\begin{align}
    P(z,C)
    =&
    \begin{cases}
    1-CmG_{II}(z)
    & \text{constant interaction}, \\
    1-C(z-M)G_{II}(z)
    & \text{WT interaction},
    \end{cases}
    \nonumber
\end{align}
we obtain the pole position $z_{R}$ by solving
\begin{align}
    P(z_{R},C)=0,
    \label{eq:polecondApp}
\end{align}
for a given $C$. Namely, the zero of $P(z,C)$ corresponds to the pole position of the scattering amplitude. Clearly, in both cases, $P(z,C)$ is an analytic function of $z$ except for the unitary cut on the real axis. It is also analytic for $C$. Suppose that we find a pole $z_{0}$ away from the unitary cut with the coupling strength $C_{0}$. Expanding the function $P(z,C)$ around the zero, we obtain
\begin{align}
    P(z,C)
    =&\sum_{n,m}a_{nm}(C-C_{0})^{n}(z-z_{0})^{m} 
    \nonumber \\
    =&
    a_{01}(z-z_{0})
    +a_{10}(C-C_{0}) \nonumber \\
    &+\mathcal{O}\bigl(\Delta_{z}^{2},\Delta_{z}\Delta_{C},\Delta_{C}^{2}\bigr)
    \nonumber
\end{align}
where $\Delta_{z}=z-z_{0}$ and $\Delta_{C}=C-C_{0}$. Thus, $P(z_{R},C)=0$ leads to 
\begin{align}
    z_{R}
    &= 
    z_{0}
    -\frac{a_{10}}{a_{01}}\Delta_{C}
    +\mathcal{O}\bigl(\Delta_{z}^{2},\Delta_{z}\Delta_{C},\Delta_{C}^{2}\bigr)
    \nonumber 
\end{align}
In the limit $C\to C_{0}$, the solution approaches $z_{R}\to z_{0}$. Thus, the pole position $z_{R}$ is a continuous function of the coupling strength $C$ (see also Ref.~\cite{Taylor} for the potential scattering.).

\begin{figure}[tbp]
\includegraphics[width=7cm,clip]{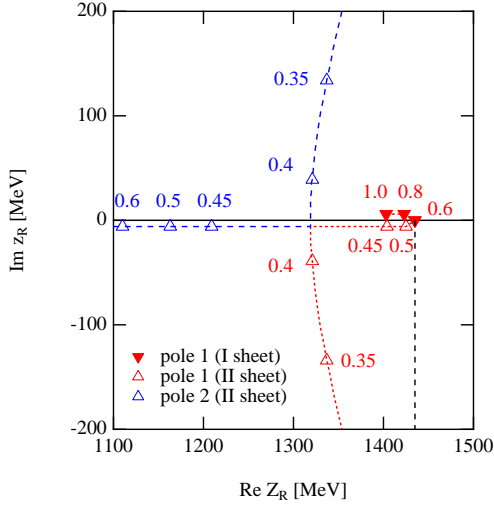}
\caption{\label{fig:pole} 
(Color online) Pole trajectories with $M=939$ MeV, $m=496$ MeV, and the natural subtraction constant. Pole 1 is the most relevant to the amplitude above the threshold. The poles on the real axis are slightly shifted to positive (negative) direction of the imaginary axis when they are on the first (second) Riemann sheet.
}
\end{figure}

For illustration, we show the pole trajectories of the scattering amplitude with the WT interaction in Fig.~\ref{fig:pole}. The masses are chosen to be $M=939$ MeV and $m=496$ MeV, and we use the natural subtraction constant $a=-1.15$. The coupling strength is given by $C=x[-4/(2f^{2})]$ with $f=92.4$ MeV, where we vary the parameter $x$ and look for the poles of the amplitude. With $x=1.0$, we obtain a bound state in the first Riemann sheet. As we decrease the interaction strength, the binding energy decreases and eventually the pole reaches the threshold at $x\sim 0.6$, which is called zero energy resonance. When the interaction strength is further reduced, the pole moves to the second Riemann sheet and becomes a virtual state. This pole goes to the lower energy direction. There is another pole which comes up from the lower side. These two poles correspond to the solution of a quadratic equation, and the trajectory of two poles matches when Eq.~\eqref{eq:polecondApp} has a double root. This meeting point is determined by the condition~\eqref{eq:meet}. Two poles then move to the complex plane, which are conjugate to each other on the second Riemann sheet. Note that the pole with the negative imaginary part is relevant to the physical scattering on the real axis. This pole  can move above the threshold where it is interpreted as a resonance. 

\section{COUPLED CHANNEL SCATTERING}\label{sec:coupledchannel}

\subsection{Chiral unitary approach in coupled-channel scattering}

Here we summarize the framework of the chiral unitary approach in the coupled-channel scattering. The Weinberg-Tomozawa (WT) energy-dependent interaction is given by
\begin{align}
    V^{\text{(WT)}}_{ij}(W)
    & =
    C_{ij}(2W-M_i-M_j)  ,
    \nonumber 
\end{align}
where $i,j$ denote the channel indices. The corresponding energy-independent interaction reads
\begin{align}
    V^{\text{(indep)}}_{ij}
    =&C_{ij}(m_i+m_j)
    \nonumber  \\
    =&V^{\text{(WT)}}_{ij}
    \left(W=\frac{M_i+m_i+M_j+m_j}{2}\right)
    . \nonumber 
\end{align}
Namely, two interactions agree at the threshold (average of the threshold energies) in the diagonal (off-diagonal) channels. The scattering amplitude is given in a matrix form as
\begin{align}
    T(W)
    &=
    [\bm{1}-V(W)G(W;a)]^{-1}V(W) ,
    \nonumber 
\end{align}
with $V_{ij}(W)$ being $V^{\text{(WT)}}_{ij}(W)$ or $V^{\text{(indep)}}_{ij}$. The loop function $G_i(W;a)$ is diagonal and given by the same formula with Eqs.~\eqref{eq:Gfn} and \eqref{eq:barq} with adding the channel index $i$. The definition of the second Riemann sheet is also same with Eqs.~\eqref{eq:amplitudeII} and \eqref{eq:GfnII}, and the condition for the pole of the amplitude is 
\begin{align}
    \begin{cases}
    \det (\bm{1}-VG)|_{W=z_R}
    =
    0 
    &\text{(bound states)}, \\
    \det (\bm{1}-VG_{II})|_{W=z_R}
    =
    0
    &\text{(virtual states/resonances)},
    \end{cases} 
    \nonumber
\end{align}
where $z_R$ is the energy of the bound state, virtual state, or resonance. The pole residue is defined as
\begin{align}
    T_{ij}(W)
    =
    \frac{g_ig_j}{W-z_R}
    +\sum_{n}a^{(n)}_{ij}(W-z_R)^n .
    \nonumber
\end{align}
In contrast to the single-channel case, the residues cannot be written analytically. As shown in Ref.~\cite{Sekihara:2010uz}, the residues $g_{i}g_{j}$ should satisfy the relation
\begin{align}
    \sum_{i,j}g_ig_j
    \left.\left(
    \frac{\partial G_{i}}{\partial W}
    \delta_{ij}
    +G_{i}
    \frac{\partial V_{ij}}{\partial W}
    G_{j}
    \right)\right|_{W\to z_{R}}
    = -1 \label{eq:ggcondcoupled} ,
\end{align}
which is the generalization of the Ward identity for the vertex-renormalization factor and the wave function renormalization factor. For the coupled-channel scattering, $\{g_{i}\}$ cannot be uniquely determined by one constraint of Eq.~\eqref{eq:ggcondcoupled}. For the single-channel case, Eq.~\eqref{eq:ggcondcoupled} leads to Eqs.~\eqref{eq:WTcoupling} and \eqref{eq:couplingWT} with the help of the pole condition~\eqref{eq:polecond}:
\begin{align}
    g^{2}
    &= -\frac{1}
    {G^{\prime}(z_{R})+G(z_{R})V^{\prime}(z_{R})G(z_{R})} 
    \nonumber \\
    &= -\frac{1}
    {G^{\prime}(z_{R})
    -\frac{V^{\prime}(z_{R})G(z_{R})}{V(z_{R})}}  ,
    \label{eq:generalg2}
\end{align}
with properly chosen Riemann sheet of the loop function. 

\subsection{Compositeness in coupled-channel scattering with multiple bound states}
\label{subsec:Zcoupledbound}

In Sec.~\ref{subsec:nonrel}, we consider the system with one bound state and the continuum states in one channel. Here we consider several elementary bound states labeled by $n$ in the coupled-channel scattering with channel index $i$. The eigenstates of the free Hamiltonian are
\begin{align}
    H_0\ket{\bm{q},i}
    & = E_{i}(\bm{q})\ket{\bm{q},i}
    = \left[
    \frac{\bm{q}^{2}}{2\mu_{i}}+E_{\text{th},i}
    \right]\ket{\bm{q},i},
    \nonumber \\
    H_0\ket{B_{n,0}}
    & = -B_{n,0}\ket{B_{n,0}}  ,
    \nonumber
\end{align}
where $E_{\text{th},i}$ is the appropriate threshold energy in channel $i$ measured from the reference channel. These eigenstates satisfy the orthonormal conditions 
\begin{align}
    \bra{\bm{q},i}\kket{\bm{q}^{\prime},j}
    &=\delta(\bm{q}^{\prime}-\bm{q})\delta_{ij},
    \nonumber 
    \\
    \bra{B_{n,0}}\kket{B_{m,0}}
    &=\delta_{nm}, 
    \quad \bra{B_{n,0}}\kket{\bm{q},i}
    =0 
    \nonumber .
\end{align}
The complete set is expanded as
\begin{align}
    1 
    &= \sum_{n}\ket{B_{n,0}}\bra{B_{n,0}} 
    + \sum_i\int d\bm{q} \ket{\bm{q},i}\bra{\bm{q},i} .
    \nonumber
\end{align}
Let us label the bound state in the full Hamiltonian by $\alpha$. In general, the number of $\alpha$ is not necessary the same with the number of $n$. Now we define $Z_{n}^{\alpha}$ and $X_i^{\alpha}$ for each bound state $\alpha$ as
\begin{align*}
    Z_{n}^{\alpha}
    &=|\bra{B_{n,0}}\kket{B^{\alpha}}|^2,
    \quad
    X_i^{\alpha}
    =\int d\bm{q}|\bra{\bm{q},i}\kket{B^{\alpha}}|^2, \\
    1 
    &= \sum_{n}Z_{n}^{\alpha} + \sum_{i} X_i^{\alpha}
    \quad \text{for each } \alpha ,
\end{align*}
where $X_i^{\alpha}$ ($Z_{n}^{\alpha}$) represents the probability of finding the bound state $B^{\alpha}$ in the scattering state in channel $i$ (elementary state $B_{n,0}$). If we regard the $\{B_{n,0}\}$ states as elementary particles, the compositeness of the bound state $B^{\alpha}$ is given by
\begin{align*}
    X^{\alpha}
    &=\sum_{i} X_i^{\alpha}
    =1 - \sum_{n}Z_{n}^{\alpha}
\end{align*}
with $X_i^{\alpha}$ being the components in channel $i$. 

With the Schr\"odinger equation, we obtain
\begin{align}
    X^{\alpha}
    &= 
    \sum_{i}
    \int d\bm{q}\frac{|\bra{\bm{q},i}V\ket{B^{\alpha}}|^2}{[E_i(\bm{q})+B^{\alpha}]^2} .
    \nonumber
\end{align}
For the $s$-wave scattering, defining $\bra{\bm{q},i}V\ket{B^{\alpha}}= G^{\alpha}_{W,i}[E_{i}(\bm{q})]$, we have 
\begin{align*}
    X^{\alpha}
    &= 
    \sum_{i}
    4\pi\sqrt{2\mu_{i}^{3}}
    \int_{E_{\text{th},i}}^{\infty} dE
    \sqrt{E-E_{\text{th},i}}
    \frac{|G^{\alpha}_{W,i}(E)|^{2}}
    {(E+B^{\alpha})^2} .
\end{align*}
as an analogy with Eq.~\eqref{eq:compositenessNR}. Noting that the scattering amplitude can be written as 
\begin{align*}
    t_{ij}(E)
    = &
    v_{ij}
    +
    \sum_{\alpha}
    \frac{G^{\alpha}_{W,i}(E)G^{\alpha}_{W,j}(E)}
    {E+B^{\alpha}} \\
    +&
    \sum_{k}
    4\pi\sqrt{2\mu_{k}^{3}}
    \int_{E_{\text{th},k}}^{\infty} dE^{\prime}
    \sqrt{E^{\prime}-E_{\text{th},k}}
    \frac{t_{ik}(E^{\prime})t_{kj}(E^{\prime})}
    {E-E^{\prime}+i\epsilon} ,
\end{align*}
we can express the compositeness as 
\begin{align*}
    X^{\alpha}
    =& 
    \sum_{i}
    4\pi\sqrt{2\mu_{i}^{3}}
    \int_{E_{\text{th},i}}^{\infty} dE
    \frac{\sqrt{E-E_{\text{th},i}}}
    {E+B^{\alpha}}  \\
    &\times 
    \Biggl[
    t_{ii}(E)
    -\sum_{\beta\neq\alpha}
    \frac{|G^{\beta}_{W,i}(E)|^2}{E+B^{\beta}}
    -v_{ii} \\
    &-\sum_{k}
    4\pi\sqrt{2\mu_{k}^{3}}
    \int_{E_{\text{th},k}}^{\infty} dE^{\prime}
    \sqrt{E^{\prime}-E_{\text{th},k}}
    \frac{|t_{ik}(E^{\prime})|^{2}}
    {E-E^{\prime}+i\epsilon}
    \Biggr].
\end{align*}
This is the generalization of Eq.~\eqref{eq:compositenessNRexact}. If there is only one bound state, the second term in the parentheses does not appear and the compositeness of the bound state can be written in terms of the amplitude $t$ and the interaction $v$. Using pole dominance, we obtain 
\begin{align*}
    X^{\alpha}
    \approx
    & 
    \sum_{i}
    4\pi\sqrt{2\mu_{i}^{3}}
    \frac{(g^{\alpha}_{W,i})^{2}}{\sqrt{E_{\text{th},i}+B^{\alpha}}}
    ,
\end{align*}
where we define $g^{\alpha}_{W,i}\equiv G^{\alpha}_{W,i}(E=-B^{\alpha})$.

\section{COMPOSITENESS OF THE SCALAR MESON IN THE PSEUDOSCALAR MESON SCATTERING}
\label{sec:scalar}

Here we briefly derive the expression of the compositeness in a different scattering system. We consider the field theory with a pseudoscalar meson field $\phi(0^{-})$ and a bare field of scalar meson $\Phi_0(0^{+})$ as
\begin{align}
    \mathcal{L}_0
    =&
    \tfrac{1}{2}(\partial_{\mu}\phi \partial^{\mu}\phi 
    -m^2\phi^2)
    +\tfrac{1}{2}(\partial_{\mu}\Phi_{0} \partial^{\mu}\Phi_{0} 
    -M_{\Phi_{0}}^2\Phi_{0}^2) ,
    \nonumber
\end{align}
which interact through the three-point vertex,
\begin{align}
    \mathcal{L}_{\text{int}}
    &=
    \lambda_0\Phi_{0}\phi^{2} ,
    \nonumber
\end{align}
where $\lambda_0$ is the bare coupling constant. We then consider that the spectrum of the full theory has a scalar meson bound state with mass $M_{\Phi}$. The definition of the field renormalization constant, which we denote $Z$, is the residue of the full Green's function at $s=M_{\Phi}^{2}$, namely,
\begin{align}
    \Delta(s)
    &=
    \frac{Z}{s-M_{\Phi}^{2}} ,
    \nonumber
\end{align}
where $s$ is the total momentum squared. 

Following the same argument with Sec.~\ref{subsec:rel}, we arrive at the expression
\begin{align}
    Z
    &= \frac{1}{1-\lambda_0^2\bar{G}^{\prime}(M_{\Phi}^{2})}
    \nonumber ,
\end{align}
where
\begin{align}
    \bar{G}(s)
    &= i\int\frac{d^{4}q}{(2\pi)^{4}}
    \frac{1}{(P-q)^{2}-m^{2}+i\epsilon}
    \frac{1}{q^{2}-m^{2}+i\epsilon}, 
    \nonumber \\
    \bar{G}^{\prime}(M_{\Phi}^{2})
    &= \left.
    \frac{\partial \bar{G}(s)}{\partial s}
    \right|_{s=M_{\Phi}^{2}}
    \nonumber ,
\end{align}
with $s=\sqrt{P^{2}}$. Defining the physical coupling constant $\lambda$ as the residue of the bound state pole in the scattering amplitude, we can write the compositeness as 
\begin{align}
    1-Z
    &=
    -\lambda^2\bar{G}^{\prime}(M_{\Phi}^{2}) \nonumber .
\end{align}
Thus, we again find that the compositeness is proportional to the physical coupling constant squared and the derivative of the loop function at the mass of the bound state.


\end{document}